\documentclass[a4paper,12pt]{article}

\pdfoutput=1
\usepackage{jheppub}
\usepackage[utf8]{inputenc}
\usepackage{nth}
\usepackage{physics}
\usepackage{appendix}
\usepackage{subcaption}
\usepackage{float}
\allowdisplaybreaks

\newcommand{\sumtodeg}{\ensuremath{\sum_{i_{1}, i_{2}, ..., i_{\deg}=0}^{N-1}}}

\newcommand{\dnx}[2]{\ensuremath{\text{d}^{#1}{#2}}}
\newcommand{\dx}[1]{\ensuremath{\text{d}{#1}}}
\newcommand{\Mfive}{\ensuremath{M_{(5)}^3}}
\newcommand{\Mfour}{\ensuremath{M_{(4)}^2}}

\newcommand{\bphi}{\ensuremath{\boldsymbol\phi}}

\newcommand{\hodge}{{\star}}

\newcommand{\vbein}[3]{\ensuremath{e_{#2}^{({#1}){#3}}}}
\newcommand{\ivbein}[3]{\ensuremath{e_{\;\;\;\;\;{#3}}^{({#1}){#2}}}}

\newcommand{\Tcoeffs}{\ensuremath{T_{i_{1} i_{2} ... i_{\deg}}}}

\newcommand{\Tz}{\ensuremath{T_{iiii}}}
\newcommand{\Ti}{\ensuremath{T_{iii,i+1}}}
\newcommand{\Tii}{\ensuremath{T_{ii,i+1,i+1}}}
\newcommand{\Tiii}{\ensuremath{T_{i,i+1,i+1,i+1}}}

\newcommand{\ti}{\ensuremath{T_{iii,i-1}}}
\newcommand{\tii}{\ensuremath{T_{ii,i-1,i-1}}}
\newcommand{\tiii}{\ensuremath{T_{i,i-1,i-1,i-1}}}

\numberwithin{equation}{section}

\title{\boldmath Clockwork Cosmology}
\author[1,2]{Kieran Wood,}
\author[1,2]{Paul M. Saffin,}
\author[1,2]{and Anastasios Avgoustidis}
\affiliation[1]{School of Physics and Astronomy, University Of Nottingham,\\ Nottingham NG7 2RD, UK}
\affiliation[2]{Nottingham Centre Of Gravity, \\Nottingham NG7 2RD, UK}
\emailAdd{kieran.wood@nottingham.ac.uk}
\emailAdd{paul.saffin@nottingham.ac.uk}
\emailAdd{anastasios.avgoustidis@nottingham.ac.uk}

\abstract{The higher order generalisation of the clockwork mechanism to gravitational interactions provides a means to generate an exponentially suppressed coupling to matter from a fundamental theory of multiple interacting gravitons, without introducing large hierarchies in the underlying potential and without the need for a dilaton, suggesting a possible application to the hierarchy problem. We work in the framework of ghost free multi-gravity with ``nearest-neighbour'' interactions, and present a formalism by which one is able to construct potentials such that the theory will always exhibit this clockwork effect. We also consider cosmological solutions to the general theory, where all metrics are of FRW form, with site-dependent scale factors/lapses. We demonstrate the existence of multiple deSitter vacua where all metrics share the same Hubble parameter, and we solve the modified Einstein equations numerically for an example clockwork model constructed using our formalism, finding that the evolution of the metric that matter couples to is essentially equivalent to that of general relativity at the modified Planck scale. It is important to stress that while we focus on the application to clockwork theories, our work is entirely general and facilitates finding cosmological solutions to any ghost free multi-gravity theory with ``nearest-neighbour" interactions. Moreover, we clarify previous work on the continuum limit of the theory, which is generically a scalar-tensor braneworld, using the Randall-Sundrum model as a special case and showing how the discrete-clockwork cosmological results map to the continuum results in the appropriate limit.}

\begin{document}

\maketitle
\flushbottom

\section{Introduction}\label{Sec:Intro}

The existence of an exponentially large hierarchy between the interaction scales of electroweak and gravitational physics remains somewhat shrouded in mystery. Generically, the mass of the Higgs boson is quadratically unstable to radiative corrections arising from any new physics with a mass scale in this large UV window; this is the \emph{hierarchy problem}. Over the years, myriad potential solutions to this problem have been proposed, though crucially no experimental evidence that favours any one particular model has been forthcoming  \cite{Beyond_SM_lectures,Criteria_for_natural_hierarchies}, and so it remains important to consider new ideas. Solutions are often based on supersymmetry \cite{Supergravity}, where loop cancellations protect the Higgs mass from such corrections. However, there is another school of thought who purport that the apparent exponential hierarchy we observe is deceptive, and that the mass scale of gravity is actually much closer to the electroweak scale. Historically, this deception has been attributed to (e.g.) the presence of a warped \cite{RS1} or large \cite{Hierarchy_at_mm} extra dimension, but more recently a new set of \emph{clockwork} models \cite{ClockworkGrav,Deconstructing} have emerged, which show signs of promise.

Underpinning the clockwork ethos is the realisation that the usual identification one would make between new physics effects (i.e. UV completion) and their corresponding interaction scales is not not necessarily a correct one -- a hidden assumption is present. Interaction scales characterise the strength of some effective interaction, whereas UV completion refers to the mass scale at which new degrees of freedom must enter, and these two quantities are \emph{incommensurable}. If one were to take some arbitrary Lagrangian in natural units and reinsert factors of $\hbar$ and $c$, it would be immediately clear that masses and interaction scales have different dimensions, and in fact that the commensurable quantities are masses with \emph{products} of scales and couplings (see \cite{Giudice} and related discussions in \cite{hbar_expansion} for an explicit demonstration). Indeed, this is why we have been careful thus far when referring to mass scales and interaction scales, so as to emphasise their distinction. In natural units, most couplings are typically $\mathcal{O}(1)$, so the identification usually works in practice, but it could be the case that UV completion occurs at a much lower energy than the associated interaction scale, if we were to have particularly small couplings. It is in this sense that one is able to solve the hierarchy problem, by using a small enough coupling so that quantum gravity effects can enter at a mass scale small enough to not bother the Higgs, \emph{while still maintaining Planck scale interactions}. The question then is how we might obtain such a small coupling in a natural manner; this is the purpose of the clockwork.

The clockwork mechanism was initially proposed to construct an axion setup where the effective axion decay constant becomes super-Planckian \cite{Rattazzi}, as is required by cosmological relaxation models \cite{Relaxation}, but the general idea has since been generalised to a much wider class of fields \cite{Giudice}. The premise is to use a chain of pairwise-interacting fields to generate a hierarchy between the parameters of the fundamental theory and the effective coupling to an external matter source, in a manner akin to the mechanism of gears in a clock (hence the name).

\begin{figure}[h!]
        \centering
        \includegraphics[width=0.7\textwidth]{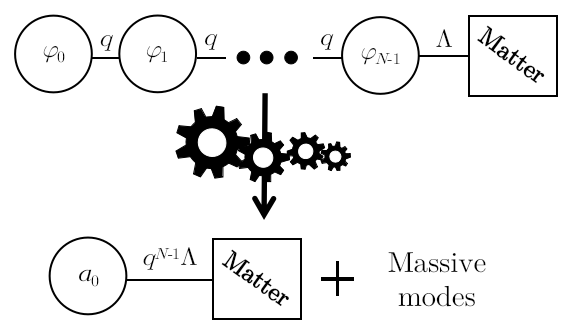}
        \caption{Schematic diagram depicting the classic clockwork setup. The gears are arranged in a 1D lattice, and we have nearest-neighbour interactions of strength $q$, which break all but one of the individual symmetries $\mathcal{S}$. The result is a zero-mode, $a_0$, associated with the unbroken subgroup $\mathcal{S}_0$, which is asymmetrically distributed through the lattice, scaling as $a_0 \sim \phi_0 + \phi_1/q + \hdots + \phi_{N-1}/q^{N-1}$. We are then able to engineer a hierarchy of scales by coupling matter to $\phi_{N-1}$, as the overlap with $a_0$ is exponentially suppressed.}
        \label{Fig:Clockwork Schematic Diagram}
    \end{figure}

The framework is as follows \cite{Giudice}: suppose we have a system which contains $N$ fields, $\phi_i$, referred to as `gears' (following the analogy), arranged in a 1D-lattice in theory space. The gears describe $N$ particles, which remain massless due to $N$ copies of some symmetry $\mathcal{S}$. Neither the explicit nature of the fields nor that of the symmetries are important, but we know that the full symmetry group of the theory contains at least the product $\mathcal{S}^{N}$ . Now suppose we introduce an interaction potential between the fields linking nearest-neighbours, characterised by some parameter $q>1$ which treats the sites \emph{asymmetrically}. Since we are working on a 1D-lattice with boundaries, we have only $N-1$ interactions, and since each interaction breaks only the symmetry corresponding to each individual site, one diagonal copy of $\mathcal{S}$ survives the breaking. As a result, the system possesses a massless zero-mode i.e. some linear combination of the original $\phi_i$ fields which has mass eigenvalue 0, as well as a tower of massive modes on top. This zero-mode is, however, not uniformly distributed throughout the lattice. Because the interaction treats the sites asymmetrically, the distribution of the zero-mode throughout the lattice is also asymmetric, and becomes exponentially suppressed at one end. Thus, by coupling some matter fields to the gear at the suppressed end of the lattice, one can engineer an exponentially suppressed coupling to the zero-mode. This idea has seen various interesting applications in recent years \cite{Clockwork_WIMP,Flavour_puzzle,Clockwork_inflation,Clockwork_goldstones}, and has since been generalised to allow also for non-nearest-neighbour interactions \cite{Non-nearest_neighbour}. A schematic diagram of the classic setup is shown in Fig. \ref{Fig:Clockwork Schematic Diagram}.

If we apply the clockwork philosophy to gravitational physics, taking our clockwork gears to be gravitons and our symmetries to be $N$ copies of diffeomorphism invariance, then we naively have a solution to the aforementioned hierarchy problem, with the clockwork interactions breaking the overall symmetry down to one asymmetrically distributed diagonal subgroup of diffeomorphisms associated with the surviving massless graviton. 

As ever with gravity, things are not quite so simple. In \cite{Disassembling_the_clockwork}, it was demonstrated that one cannot apply the clockwork mechanism in the traditional sense \cite{Giudice} to non-Abelian theories, by using a series of elegant group-theoretic arguments to show that such an asymmetric structure in the unbroken diagonal subgroup is forbidden. The argument rested upon the assumption of a lack of site dependence in the couplings between gears, as was the case in the original proposal. The authors of \cite{Giudice} retorted in \cite{Giudice_rebuttal}, saying that such site independence could be little more than a statement about the full UV model, if one views the clockwork as a low-energy EFT. Regardless, if we allow for site dependence in the couplings then we can once again obtain interesting phenomenology, though there is some disagreement about whether this defeats the object \cite{Disassembling_the_clockwork}, as including site dependence necessarily means that we must have a degree of hierarchy in the underlying fundamental theory. Such hierarchies in the fundamental theory have in the past been accounted for by introducing a dilaton \cite{Dilaton}; here we instead take the view that a small hierarchy in the fundamental theory can be acceptable so long as the couplings $q$ remain roughly of order unity.

With this in mind, more opportunities avail themselves.  In \cite{ClockworkGrav}, it was shown that the desired asymmetric distribution of the graviton zero-mode could be obtained in an entirely new manner through a higher order generalisation of the standard clockwork mechanism, from a fundamental theory with only hierarchies up to $q^2$, and so no large parameters, provided that $q\sim\mathcal{O}(1)$. The unbroken diagonal subgroup of diffeomorphisms is symmetric, as it must be \cite{Disassembling_the_clockwork}, but the underlying \emph{background} vacua follow an asymmetric distribution -- the action of the symmetric subgroup on their fluctuations then results in the desired zero-mode distribution. This situation was not considered in \cite{Disassembling_the_clockwork}, whose background was Minkowski at all sites, and it offers us a very enticing new prospect to solve the hierarchy problem.  Later, it was shown in \cite{Deconstructing}, following the dimensional deconstruction philosophy of \cite{Discrete_gravitational_dimensions}, that the general ghost free multi-gravity action used in \cite{ClockworkGrav,Interacting_spin_2} can be viewed as the deconstruction of a general 5-dimensional continuum theory, where the discrete set of gears are considered simply as the induced metrics on 4D hypersurfaces at discrete locations along the extra dimension. In particular, it was shown that within the class of continuum gravitational clockwork theories reside scalar-tensor theories where the scalar should \emph{not} be identified with a dilaton, distinguishing them from previous approaches. The continuum picture naturally has the interpretation of a braneworld model, where coupling matter to one of the gears becomes analogous to placing matter on a brane at the corresponding location in the \nth{5} dimension. It is no surprise, then, that amongst the special cases of the general theory put forward in \cite{Deconstructing} resides the quintessential braneworld model of Randall and Sundrum, RS1 \cite{RS1} (though without any $\mathbb{Z}_2$ symmetry -- we will discuss this in Sections \ref{Sec:Continuum} and \ref{Sec:RS}). 

Clearly, much progress has been made surrounding the application of the clockwork mechanism to gravity in recent years, but still there is much to be discovered. Indeed, we do not actually know much about which explicit forms of interaction potential are able to produce the asymmetric background vacuum structure needed to clockwork the graviton gears, besides the existence of the particular case considered in \cite{ClockworkGrav}. Even less is known about the cosmology of such models, though one might expect some clockworks to possess interesting phenomenology worth studying, given the presence of RS1 as a special case in the continuum version of the theory. The aim of the present work is to shed light on both of these issues, as we develop a general formalism utilising symmetric polynomials to determine which interactions give a valid clockwork vacuum, and begin to look at the cosmology by determining the background evolution for two example models. One of these is essentially a deconstruction of RS1, which serves as a good consistency check for our formalism as the phenomenological consequences are well-studied in the continuum limit \cite{Carsten_review,Langlois_review}. The formalism we present is entirely general and provides a means to determine the background cosmological solutions to \emph{any} modified theory of gravity utilising multiple pairwise interacting metric fields, although in the present work we specialise to clockwork gravity models due to their potential application to the hierarchy problem.

The structure of the paper, then, is as follows: in Section \ref{Sec:Formalism} we outline the specifics of clockwork theory and develop our symmetric polynomial formalism; in Section \ref{Sec:Discrete} we specialise to clockwork gravity; in Section \ref{Sec:Continuum} we show how the continuum theory arises and explain how the construction of the \nth{5} dimension must differ from the usual RS1 orbifold; in Section \ref{Sec:Examples} we look at the background cosmology of two example models, and finally we conclude in Section \ref{Sec:Conclusion}.

Throughout, we work in natural units $\hbar=c=1$ and use a mostly-plus metric signature.

\section{The Clockwork Mechanism}\label{Sec:Formalism}

In order to implement the clockwork mechanism as outlined above, we need an action for a chain of $N$ fields originally exhibiting copies of some symmetry $\mathcal{S}$, which will be broken by introducing interactions. Of course, the ultimate goal here is to choose our fields to be gravitons and our symmetries to be diffeomorphisms, but in this Section we shall work only with scalar fields, whose associated symmetries are Goldstone shifts inherited from $N$ copies of $U(1)$. We do this for two reasons: first, for clarity -- the formalism we introduce is much more intuitive in the language of scalars, and second, as we will see later on, the clockwork gravity potential is identical to the scalar potential when we look for the background vacuum solution, with the respective conformal factors in the metrics (which we said fully determine the structure of the graviton zero-mode \cite{ClockworkGrav}) playing the role of the scalar fields.

The Lagrangian we choose is one for $N$ real scalar fields $\boldsymbol\phi = (\phi_0, \phi_1, ... , \phi_{N-1})$, which interact via some homogeneous polynomial potential of degree ``$\deg$", in $D$ dimensions:
\begin{align}\label{action}
    S &= \int d^D x \, \mathcal{L}(\boldsymbol\phi , \partial \boldsymbol\phi)
    \\
    \label{Lagrangian}
    \mathcal{L}(\boldsymbol\phi , \partial \boldsymbol\phi) &= -\frac12 \sum_{i=0}^{N-1} \partial_\mu \phi_i \partial^\mu \phi_i - V(\boldsymbol\phi) \, ,
\end{align}
where the potential is given by:
\begin{equation} \label{Potential}
    V(\boldsymbol\phi) = \sumtodeg \Tcoeffs \phi_{i_1} \phi_{i_2} ... \phi_{i_{\deg}} \, ,
\end{equation}
and the coefficients $\Tcoeffs=T_{(i_{1} i_{2} ... i_{\deg})}$ are totally symmetric. The action has a scaling symmetry when we take $\deg=2D/(D-2)$. Although the clockwork has been generalised to allow for non-nearest-neighbour interactions \cite{Non-nearest_neighbour}, we will restrict ourselves to nearest-neighbour anyway in the knowledge that, when working with gravity, non-nearest neighbour (loop-type) interactions generically lead to undesirable Boulware-Deser ghosts \cite{BD_Ghosts,GhostFreedom,Cycles}. This means that our coefficients \Tcoeffs are restricted to only terms of the form $T_{iii...}$, $T_{i+1,ii...}$, $T_{i+1,i+1,i...}$ and so on, i.e. the potential contains only terms that look like $\phi_0^4$, $\phi_0^3\phi_1$, $\phi_0^2\phi_1^2$ etc.

The equations of motion that result from the Lagrangian Eq. \eqref{Lagrangian} are:
\begin{equation} \label{eom}
    \partial_\mu \partial^\mu \phi_a - \deg \sum_{i_{2}, i_{3}, ..., i_{\deg}=0}^{N-1} T_{a i_2 i_3 ... i_{\deg}} \phi_{i_2} \phi_{i_3} ... \phi_{i_{\deg}}= 0 \, ,
\end{equation}
where the factor $\deg$ arises as a result of the symmetry in the coefficients. If we want there to exist a vacuum solution at $\bphi = \textbf{c} =  (c_0, c_1, ... , c_{\deg})$, we require:
\begin{equation}\label{VacuumCondition}
    \sum_{i_{2}, i_{3}, ..., i_{\deg}=0}^{N-1} T_{a i_2 i_3 ... i_{\deg}} c_{i_2} c_{i_3} ... c_{i_{\deg}}= 0 \, .
\end{equation}
Perturbing the vacuum solution, $\bphi = \textbf{c} + \delta\bphi$, the fluctuations have dynamics determined via the following second-order action:
\begin{equation}\label{Action2}
    S^{(2)} = \int d^D x \, \left[ -\frac12 \sum_{i=0}^{N-1} \partial_\mu \delta\phi_i \partial_\mu \delta\phi_i - \frac12 \deg (\deg-1) \sum_{i_{3}, i_{4}, ..., i_{\deg}=0}^{N-1} T_{a b i_3 ... i_{\deg}} c_{i_3} c_{i_4} ... c_{i_{\deg}} \delta\phi_a \delta\phi_b \right] \, ,
\end{equation}
giving the mass matrix:
\begin{equation}\label{MassMatrix}
    M^2_{ab} = \deg(\deg-1) \sum_{i_{3}, i_{4}, ..., i_{\deg}=0}^{N-1} T_{a b i_3 ... i_{\deg}} c_{i_3} c_{i_4} ... c_{i_{\deg}} \, .
\end{equation}

Immediately we can see the presence of the zero-mode from equations \eqref{VacuumCondition} and \eqref{MassMatrix},
\begin{equation}\label{ZeroMode}
    \sum_{b=0}^{N-1} M^2_{ab} c_b = 0 \, ,
\end{equation}
and hence there is a flat direction along transformations for which $\delta\phi_a \propto c_a$.

 We would like to impose a hierarchy on the vacuum structure such that one end of the chain of fields is exponentially suppressed compared to the other; this is the defining feature of a clockwork model. That is, we want to take our vacuum solution $\bphi=\mathbf{c}$ to look something like:
\begin{equation}\label{VacStructure}
    c_a = \frac{c}{q^a} \, ,
\end{equation}
for some constant $q \gtrapprox1$ and some universal scale $c$. This choice of vacuum is not unique; in principle any vacuum possessing a hierarchy with $c_{i+1}/c_i < 1$ would suffice. Eq. \eqref{VacStructure} is simply a natural choice and serves well to develop our formalism. The question we then want to answer is whether we can determine a set of symmetric coefficients $\Tcoeffs$ in the potential Eq. \eqref{Potential} such that the vacuum has this hierarchy but the coefficients themselves do not. To begin to answer this question, we turn to symmetric polynomials to reformulate the problem.

\subsection{Symmetric Polynomial Formalism}\label{Sec:SymPols}

We start this section by defining objects called the \emph{elementary symmetric polynomials}:
\begin{equation}\label{ElSymPols}
    \begin{split}
        e_0(x_1,x_2,...,x_{\deg}) &= 1 \\
        e_1(x_1,x_2,...,x_{\deg}) &= \sum_{1\leq i \leq \deg} x_i \\
        e_2(x_1,x_2,...,x_{\deg}) &= \sum_{1\leq i<j\leq\deg} x_i x_j \\
        \vdots \\
        e_k(x_1,x_2,...,x_{\deg}) &= \sum_{1\leq j_1<j_2<...<j_k\leq \deg} x_{j_1}...x_{j_k} \\
        \vdots \\
        e_{\deg}(x_1,x_2,...,x_{\deg}) &= x_1 x_2 ... x_{\deg} \, .
    \end{split}
\end{equation}
Each $e_k$ contains $\binom{\deg}{k}=\frac{\deg!}{k!(\deg-k)!}$ terms in total. The elementary symmetric polynomials are special because \emph{any} symmetric polynomial of degree $\deg$ can be expressed in terms of sums and products of the elementary symmetric polynomials up to $e_{\deg}$; in essence they act as a basis for general symmetric polynomials (hence their name).

The $e_k$ are useful because we know that our coefficients \Tcoeffs in the potential are totally symmetric, so we can introduce the potential polynomial,
\begin{equation}\label{PotPoly}
    T(x_1,...,x_{\deg}) = \sumtodeg \Tcoeffs (x_1)^{i_1}(x_2)^{i_2}...(x_{\deg})^{i_{\deg}} \, ,
\end{equation}
which is then manifestly symmetric in all of its arguments, and hence can be expressed in terms of the elementary symmetric polynomials Eqs. \eqref{ElSymPols}. Then, when combined with Eqs. \eqref{VacuumCondition} and \eqref{VacStructure}, we get the requirement on the potential polynomial, and hence on \Tcoeffs, that will give us the desired vacuum structure:
\begin{equation}\label{VacConditionPolynomial}
    T(x_1,q^{-1},q^{-1},...,q^{-1}) = 0 \, .
\end{equation}
In addition to this condition, we want $T$ to contain only low powers of $q$ so that there is no fundamental hierarchy in the coefficients \Tcoeffs.

We can do a similar thing for the mass matrix, and introduce the mass polynomial,
\begin{equation}\label{MassPoly}
    \begin{split}
        M(x_1,x_2) &= \sum_{i_1,i_2=0}^{N-1} M_{i_1 i_2}^2 (x_1)^{i_1} (x_2)^{i_2}
        \\
        &= c^{\deg-2} \deg(\deg-1) \sumtodeg \Tcoeffs (x_1)^{i_1} (x_2)^{i_2} q^{-i_3} q^{-i_4}...q^{-i_{\deg}}
        \\
        &= c^{\deg-2} \deg(\deg-1) T(x_1,x_2,q^{-1},...,q^{-1}) \, ,
    \end{split}
\end{equation}
from which we should readily be able to determine the components of the mass matrix, once the coefficients \Tcoeffs have been determined.

Currently, there is a lot of freedom in choosing a potential polynomial that may do the job for us. However, our restriction to nearest-neighbour interactions constrains the form of $T$ to \emph{only} a linear combination of terms of the form
\begin{equation}\label{T_building_block}
    T \supset \kappa (e_{\deg})^p e_i \; ,
\end{equation}
for constant $\kappa$ and some power $0\leq p\leq N-1$. Given this restriction, we can construct the most general possible potential polynomial as:
\begin{equation}\label{T_nearestneighbour}
    T(x_1,...,x_{\deg}) = \sum_{n=0}^{N-2} \alpha_n \left(e_{\deg}(x_1,...,x_{\deg})\right)^n \sum_{m=0}^{\deg} \beta_m e_m(x_1,...,x_{\deg}) \, ,
\end{equation}
where $\alpha_n$ and $\beta_m$ are constants. This is nothing more than a general linear combination of terms of the form Eq. \eqref{T_building_block}. The coefficients can be read off as:
\begin{equation}\label{T_components}
    \begin{split}
        T_{pppp...p} &= \alpha_p\beta_0 + \alpha_{p-1}\beta_{\deg} \\
        T_{\{p+1\}^q\{p\}^{\deg-q}} &= \alpha_p \beta_q \, ,
    \end{split}
\end{equation}
for $p=0,...,N-1$ and $q=1,...,\deg-1$, with $\alpha_{N-1}=\alpha_{-1}=0$, and all other $\Tcoeffs=0$. Then, all we need to construct a theory with the desired vacuum hierarchy is to find a set of $\alpha_n$ and $\beta_m$ such that Eq. \eqref{VacConditionPolynomial} is satisfied, with sufficiently small powers of $q$ to avoid that same hierarchy in \Tcoeffs.

We can go further here, because we can evaluate the elementary symmetric polynomials at $(x_1, q^{-1},...,q^{-1})$ -- they are:
\begin{equation}\label{ElSymPolsQs}
    e_k(x_1, q^{-1},...,q^{-1}) = \binom{\deg-1}{k-1}x_1 q^{1-k} + \binom{\deg-1}{k}q^{-k} \; ,
\end{equation}
so, the vacuum condition Eq. \eqref{VacConditionPolynomial} reads:
\begin{equation}\label{VacCondBinom}
   \sum_{n=0}^{N-2} \alpha_n x_1^n q^{n(1-\deg)} \sum_{m=0}^{\deg} \beta_m \left[ \binom{\deg-1}{m-1}x_1 q^{1-m} + \binom{\deg-1}{m}q^{-m} \right] = 0 \, .
\end{equation}
Then, comparing coefficients either side, we see that the requirement is that the coefficients of all powers of $x_1$ must vanish separately, which leads to the following:
\begin{equation}\label{FullVacCond}
     \alpha_i \sum_{m=0}^{\deg} \beta_m \binom{\deg-1}{m}q^{-m} + q^{\deg} \alpha_{i-1}  \sum_{m=0}^{\deg} \beta_m \binom{\deg-1}{m-1}q^{-m} = 0 \;\;\; \forall i
\end{equation}
If we can find a set of $\alpha_n$ and $\beta_m$ that satisfies Eq. \eqref{FullVacCond}, then we can build a clockwork potential that provides the desired asymmetrically distributed vacuum solution\footnote{Note that here, because of the two equations at the end points i.e. $i=0$ and $i=N-1$ where one of the $\alpha$'s vanishes, both of the two sums involving the $\beta_m$'s must vanish separately. However, later in Section \ref{Sec:Discrete} we will see that including matter on the boundaries adds an extra term to Eq. \eqref{FullVacCond} that stops this from being true, so we leave the condition in full here.}. One is free to make this even simpler, by setting all of the non-zero $\alpha$'s to be equal, since this is just a choice of potential, which turns the above into a condition only on the $\beta$'s. We will indeed do this going forward to make our lives easier, but for now the $\alpha_n$ remain for completeness.

We can do a similar thing for the mass polynomial Eq. \eqref{MassPoly}, where we need to evaluate $e_k(x_1,x_2,q^{-1},...,q^{-1})$. These follow a similar pattern:
\begin{equation}\label{MassElSymPols}
    e_k(x_1, x_2, q^{-1},...,q^{-1}) = \binom{\deg-2}{k-2} x_1 x_2 q^{2-k} + \binom{\deg-2}{k-1}(x_1+x_2) q^{1-k} + \binom{\deg-2}{k}q^{-k} \, .
\end{equation}
Substituting into our mass polynomial Eq. \eqref{MassPoly} yields the following:
\begin{equation}
    M(x_1,x_2) = K \sum_{n=0}^{N-2} \alpha_n \left(e_{\deg}(x_1,x_2,q^{-1},...,q^{-1})\right)^n \sum_{m=0}^{\deg} \beta_m e_m(x_1,x_2,q^{-1},...,q^{-1}) \, ,
\end{equation}
defining the constant $K=c^{\deg-2}\deg(\deg-1)$. Expanded out in full, the mass polynomial reads:
\begin{equation}
    \begin{split}
        M(x_1,x_2) = &K \sum_{n=0}^{N-2} \alpha_n (x_1 x_2)^n q^{n(2-\deg)} 
        \\
        &\times \sum_{m=0}^{\deg}  \beta_m \left[ \binom{\deg-2}{m-2} x_1 x_2 q^{2-m} + \binom{\deg-2}{m-1}(x_1+x_2) q^{1-m} + \binom{\deg-2}{m}q^{-m} \right] \, .
    \end{split}
\end{equation}
From this expression, we can read off the components of the mass matrix,
\begin{equation}
    \begin{split}
        M_{pp}^2 &= K \left[ \alpha_p q^{p(2-\deg)} \sum_{m=0}^{\deg} \beta_m \binom{\deg-2}{m}q^{-m} + \alpha_{p-1}q^{(p-1)(2-\deg)} \sum_{m=0}^{\deg} \beta_m \binom{\deg-2}{m-2}q^{2-m} \right]
        \\
        M_{p+1,p}^2 &= M_{p,p+1}^2 = K \alpha_p q^{p(2-\deg)} \sum_{m=0}^{\deg} \beta_m \binom{\deg-2}{m-1}q^{1-m} \; ,
    \end{split}
\end{equation}
with the indices on $M^2$ running from 0 to $N-1$, again with $\alpha_{N-1}=\alpha_{-1}=0$, and all other $M^2_{i_1,i_2}=0$. We can write this in a nicer and more symmetric manner by factoring out some of the $q$'s:
\begin{equation}\label{FAT_Matrix}
\begin{split}
    M_{ij}^2 &= K (q^{\frac{2-\deg}{2}})^{i+j-1} 
    \\
    &\times \bigg\{\delta_{ij}\left[ \alpha_i q^{\frac{2-\deg}{2}} A + \alpha_{i-1} q^{-\frac{2-\deg}{2}} B \right] +\alpha_{i+j-1}(\delta_{i,j-1}+\delta_{i-1,j})C \bigg\} \, ,
\end{split}
\end{equation}
where we have defined the constants:
\begin{itemize}
\itemsep0em
    \item $A = \sum_{m=0}^{\deg} \beta_m \binom{\deg-2}{m}q^{-m}$
    \item $B = \sum_{m=0}^{\deg} \beta_m \binom{\deg-2}{m-2}q^{2-m}$
    \item $C = \sum_{m=0}^{\deg} \beta_m \binom{\deg-2}{m-1}q^{1-m}$
\end{itemize}
for brevity.

Written out in full matrix form (and taking $\alpha_n=1 \; \forall\; n\neq 0, N-1$ for simplicity), this looks like:
\begin{equation}\label{FATMatrix2}
    M^2 = K
    \begin{bmatrix}
       aQ^{-1} & C & 0 & 0 & \dots & 0 & 0
       \\
       C & (a+b)Q & CQ^2 & 0 & \dots & 0 & 0
       \\
       0 & CQ^2 & (a+b)Q^3 & CQ^4 & \dots & 0 & 0
       \\
       0 & 0 & CQ^4 & (a+b)Q^5 & \dots & 0 & 0
       \\
       \vdots & \vdots & \vdots & \vdots & \ddots & \vdots & \vdots
       \\
       0 & 0 & 0 & 0 & \dots & (a+b)Q^{2N-5} & CQ^{2N-4}
       \\
       0 & 0 & 0 & 0 & \dots & CQ^{2N-4} & bQ^{2N-3}
    \end{bmatrix}
\end{equation}
where:
\begin{itemize}
    \itemsep0em
    \item $Q = q^{\frac{2-\deg}{2}}$
    \item $a = QA$
    \item $b = Q^{-1}B$
\end{itemize}

The symmetric polynomial formalism is powerful; just by determining a set of numbers that satisfy some relatively simple condition, Eq. \eqref{FullVacCond}, we are able to determine a potential that will give the required asymmetric vacuum structure that characterises the clockwork, as well as the matrix encoding the masses of the gears. One could hope to find an analytic form for the eigenvalues of Eq. \eqref{FATMatrix2}, in order to determine the mass gap between the zero-mode and first massive mode, as this would be the first to show up in collider experiments. However, thus far this looks impossible except for when $Q=1$ i.e. when the potential is quadratic \cite{Rattazzi,Giudice,TridiagEigs}. Nevertheless, it is easy enough to calculate eigenvalues numerically on a case-by-case basis.

\subsection{Examples}

To demonstrate the usefulness of the formalism, we now use it to reproduce some results from the clockwork literature, to write down a new clockwork, and to make a statement about shift symmetric potentials with $\deg>2$.

\subsubsection{Original (quadratic) clockwork scalar}

The original clockwork proposal \cite{Rattazzi,Giudice}, as stated in Section \ref{Sec:Intro}, was an axion setup that used a $\deg=2$ shift-symmetric potential for $N$ Goldstone bosons, $\pi_i$. Here, we will start from their potential and try to obtain their mass matrix, using the techniques developed thus far (this method to find the mass matrix is perhaps a bit overkill, given that we already know the simple form of the potential, but it serves well as an illustration of the procedure).

The potential used in \cite{Rattazzi,Giudice} is the following:
\begin{equation}\label{McCullough_Pot}
     V(\pi) = \frac{m^2}{2} \sum_{j=0}^{N-1} (\pi_j - q\pi_{j+1})^2 \; .
\end{equation}
Expanding out the sum,  we can extract from this potential the coefficients \Tcoeffs:
\begin{equation}
    T_{00} = \frac{m^2}{2} \, , \;\;\; T_{(N-1)(N-1)}=\frac{m^2}{2}q^2 \, , \;\;\; T_{pp}=\frac{m^2}{2}(1+q^2) \, , \;\;\; T_{p(p+1)} = -\frac{m^2}{2}q \, ,
\end{equation}
for $p=1,...,N-2$.  We can then use Eq. \eqref{T_components} to determine the values of $\alpha_n$ and $\beta_m$ that should comprise our potential polynomial. A quick check shows that a good choice is $\alpha_n = \alpha = m^2/2$, with
\begin{equation}
    \beta_0 = 1 \, , \;\;\; \beta_1 = -q \, , \;\;\; \beta_2 = q^2 \, .
\end{equation}
In our formalism, choosing these numbers would be the starting point, and we would check that they satisfy the vacuum condition and subsequently construct the potential $V(\pi)$. With these $\alpha$'s and $\beta$'s, the potential polynomial $T(x_1,x_2)$ reads:
\begin{equation}
    T(x_1,x_2) = \frac{m^2}{2}\sum_{n=0}^{N-2}\left(e_2(x_1,x_2)\right)^n \left[1-qe_1(x_1,x_2) + q^2 e_2(x_1,x_2) \right] \, .
\end{equation}
Substitution into Eq. \eqref{FullVacCond} returns 0, with both sums vanishing separately, and so the vacuum condition is indeed satisfied as we expect.

Finally, all that remains to determine the form of the mass matrix is to calculate the constants $A, B$, $C$ and $Q$ that appear in Eq. \eqref{FATMatrix2}. This is simple in this case, since the potential is quadratic so we have $Q=1$. The remaining constants are
\begin{equation}
     A = \beta_0 \binom{0}{0}q^0 = 1
     \; , \;\;\;\;
     B = \beta_2 \binom{0}{2-2}q^{2-2} = q^2
     \; , \;\;\;\;
     C = \beta_1\binom{0}{1-1}q^{1-1} = -q \, .
\end{equation}
Substituting into Eq. \eqref{FATMatrix2} gives:
\begin{equation}\label{McCulloughMatrix}
    M = 2 \, \frac{m^2}{2}
        \begin{bmatrix}
            1 & -q & 0 & \dots & 0
            \\
            -q & 1+q^2 & -q & \dots & 0
            \\
            \vdots & \vdots & \ddots & \vdots & \vdots
            \\
            0 & 0 & \dots & 1+q^2 & -q
            \\
            0 & 0 & \dots & -q & q^2
        \end{bmatrix} \, ,
\end{equation}
which is exactly the mass matrix in \cite{Giudice}, and all is well. The authors of the original paper go on to find closed form solutions for the eigenvalues and eigenvectors of this matrix, finding in particular the presence of a zero-mode which follows exactly the structure we have set out ($c_a = c/q^a$).

\subsubsection{Quartic clockwork scalar}\label{Sec:QCS}

The second example we consider is a new clockwork theory, whose potential we build from the ground up using our basic assumptions about the coefficients \Tcoeffs. In particular, we stated that we would like to produce the desired vacuum hierarchy from a general $\deg$ potential without introducing a similar hierarchy in \Tcoeffs. Ideally, we would like this hierarchy to be only up to $q^2$, if $q\sim\mathcal{O}(1)$. In terms of the $\alpha$'s and $\beta$'s, this means that we would like to ideally have all $\alpha$'s equal and $\beta_m\propto q^{\pm 1}$ at most. 

Can we still satisfy the vacuum condition Eq. \eqref{FullVacCond} for some $\beta$'s with this property? The vacuum condition tells us that, if we assert that $\beta_m\propto q^{\pm 1}$, for \emph{any} $\deg$ only terms up to $m=2$ can be non-vanishing in order for the sums to cancel. In this case, it turns out that we can indeed satisfy Eq. \eqref{FullVacCond}, with the choice $\alpha_n =\alpha=1 \;\forall\;n$, and,
\begin{equation}
    \begin{split}
        \beta_0 &= \frac12 \deg(\deg-1)q^{-1}
        \\
        \beta_1 &= 1-\deg
        \\
        \beta_2 &= q \, ,
    \end{split}
\end{equation}
with all other $\beta_m=0$\footnote{We could also flip everything by a minus sign, and the choice would still satisfy Eq. \eqref{FullVacCond}}. The original proposal is just the $\deg=2$ version of this particular choice (with everything multiplied up by $q$); this is the generalisation to arbitrary $\deg$.

We shall consider the $\deg=4$ case as a simple albeit as yet unstudied example, and we shall come back to study the gravitational version of this theory later in Section \ref{Sec:Examples}. For $\deg=4$, we have $\beta_0=6q^{-1}, \beta_1 = -3$, and $\beta_2=q$. This means that the potential coefficients are:
\begin{itemize}
    \itemsep0em
    \item $T_{iiii} = 6q^{-1}$
    \item $T_{i+1,iii} = -3$
    \item $T_{i+1,i+1,ii} = q$
\end{itemize}
Recalling that the coefficients are symmetric on exchange of indices, we reconstruct the potential which produces the desired vacuum solution:
\begin{equation}
    \begin{split}
        V(\bphi) &= T_{0000}\phi_0^4 +4T_{1000}\phi_0^3\phi_1 + 6T_{1100}\phi_0^2\phi_1^2 + \hdots
        \\
        &= 6q^{-1}\phi_0^4 -12\phi_0^3\phi_1 + 6q\phi_0^2\phi_1^2 + \hdots
        \\
        &= \frac{6}{q}\sum_{i=0}^{N-2} \phi_i^2 (\phi_i - q\phi_{i+1})^2 \; .
    \end{split}
\end{equation}

The next step is to determine the mass matrix. For $\deg=4$ we have $Q=q^{-1}$, and for our choice of $\beta$'s the other constants are:
\begin{equation}
     A = q^{-1}
     \; , \;\;\;\;
     B = q
     \; , \;\;\;\;
     C = -1 \, ,
\end{equation}
which means that the mass matrix has components:
\begin{equation}
\begin{split}
    M_{ij}^2 &= K (q^{-1})^{i+j-1} 
    \\
    &\times \bigg\{\delta_{ij}\left[ (1-\delta_{i,N-1}) q^{-2} + (1-\delta_{i,0}) q^{2} \right] -(\delta_{i,j-1}+\delta_{i-1,j}) \bigg\} \, .
\end{split}
\end{equation}
We can diagonalise this mass matrix via an orthogonal field space transformation, $\phi_i = \sum_j O_{ij}a_j$, where the orthogonal matrix $O_{ij}$ has its columns given by the mass eigenvectors. In particular, numerical investigations show the presence of the zero-mode with the correct structure, $O_{i0}\propto q^{-i}$, and so we have used the formalism to build a valid clockwork from a $\deg=4$ potential.

We note that this mass matrix is of precisely the same form as the mass matrix derived for the orignal `higher order clockwork gravity' theory \cite{ClockworkGrav}, after a bit of massaging to get their mass matrix into the form used here (although we work with scalars, we will see in Section \ref{Sec:Discrete} that the symmetric polynomial results carry over to the static vacuum of the gravitational theory). The model of \cite{ClockworkGrav} used a rather convoluted choice for the coefficients \Tcoeffs, which consequently lead to there being a function $F(q)$ out in front of $M^2_{ij}$. This just corresponds to a more complicated choice of $\alpha_n$ for us.

\subsubsection{Shift symmetric potential}

Suppose we now have a potential of arbitrary degree which carries a shift symmetry between adjacent gears. That is, consider potentials of the form:
\begin{equation}\label{Shift_Sym_potential}
    V(\bphi) = \sum_{i=0}^{N-1} (\phi_i - q\phi_{i+1})^{\deg} \; ,
\end{equation}
which have a valley along $\phi_i = q\phi_{i+1}$. The original proposal is the $\deg=2$ case of this type of potential, but here we work with general $\deg$. 

This type of theory cannot work as a clockwork in nature: if we define the theory in terms of new fields $\chi_i=\phi_i-q\phi_{i+1}$, whose potential consists of only self-interactions of the form $\chi_i^{\deg}$, then all the gears are massless unless $\deg=2$ (when the self-interaction is a mass term). We would like to see how our formalism comes to the same conclusion.


It is not difficult to check that the set of $\beta_m$ that produces this potential is:
\begin{equation}
    \beta_m = (-q)^m \beta_0 \; ,
\end{equation}
for arbitrary $\beta_0$. Naively, this is a nice choice because it automatically satisfies the vacuum condition, due to the binomial coefficient identity,
\begin{equation}\label{binom_identity}
    \sum_{m=0}^{n} (-1)^m \binom{n}{m} = 0 \;\;\;\;\; (n>0) \; ,
\end{equation}
which forces both sums in Eq. \eqref{FullVacCond} to vanish.

However, when it comes to the mass matrix, if $\deg>2$, then the same identity forces the constants $A$, $B$ and $C$ to vanish also, and so the mass matrix is populated entirely by zeroes, as we expect\footnote{When $\deg>2$, the sums for $A,B$ and $C$ have multiple terms which always mutually cancel, but when $\deg=2$, which corresponds to the $n=0$ case of the identity \eqref{binom_identity}, the sums have only a single term, which clearly survives -- we saw this explicitly for the original scalar case.}.

\section{Discrete Clockwork Gravity}\label{Sec:Discrete}

Now we have all we need to begin to turn our attention to gravity. We stated at the start of Section \ref{Sec:Formalism} that the action for clockwork gravity is equivalent to the scalar case in vacuum, and we will see this explicitly in a short while, though some groundwork is required before then. The starting point is the standard multi-metric gravity action in the tetrad formalism (see \cite{Deconstructing,Interacting_spin_2,Cosmo_spin_2}), with $N$ Einstein-Hilbert kinetic terms and a $\deg=4$ interaction coupling the various basis 1-forms:
\begin{align}\label{CWGravityAction}
    S &= S_K + S_V + S_M
    \\
    S_K &= \sum_{i=0}^{N-1} \frac{M_{(4)i}^2}{4} \int e^{(i)a} \wedge e^{(i)b} \wedge \hodge{R^{(i)}_{ab}}\label{CWGKinetic}
    \\
    S_V &= \sum_{i,j,k,l=0}^{N-1} \int \, T_{ijkl} \varepsilon_{abcd} \, e^{(i)a} \wedge e^{(j)b} \wedge e^{(k)c} \wedge e^{(l)d}\label{CWGPotential} \; ,
\end{align}
where the $T_{ijkl}=T_{(ijkl)}$ are our symmetric coefficients from Section \ref{Sec:Formalism}, and the tetrad basis 1-forms are $e^{(i)a}=e_\mu^{(i)a} \dx{x}^\mu$, with the vierbeins defined through $g^{(i)}_{\mu\nu} = e_\mu^{(i)a} e_\nu^{(i)b} \eta_{ab}$. $S_M$ is the action for the collective matter fields coupled to the theory. Indices are raised/lowered site-wise, Latin indices with $\eta^{(i)}_{ab}$ and Greek indices with $g^{(i)}_{\mu\nu}$, while we can swap between Latin and Greek indices using the vierbeins (via change of basis).

$R^{(i)}_{ab}$ is the curvature 2-form associated with the $i$-th tetrad, with one index lowered by $\eta_{ab}$, and $\hodge{R^{(i)}_{ab}}$ is its Hodge dual (also a 2-form in $D=4$ dimensions). We have:
\begin{align}
    R^{(i)}_{ab} &= \frac12 R^{(i)}_{ab\mu\nu} \dx{x}^\mu \wedge \dx{x}^\nu
    \\
    \hodge{R^{(i)}_{ab}} &= \frac12 \sqrt{-g} R^{(i)\alpha\beta}_{\;\;\;\;\;\;\;\;ab} \varepsilon_{\alpha\beta\gamma\delta} \dx{x}^\gamma\wedge\dx{x}^\delta \; ,
\end{align}
where the $R$'s with 4 indices are components of the $i$-th Riemann tensor. This kinetic term is nothing more than the usual Einstein-Hilbert action, just written in a nicer way using differential forms, so as to make computing the equations of motion simpler -- we can see this as follows (suppressing the $(i)$ indices):
\begin{equation*}
    \begin{split}
        S_K &= \frac{M_{(4)}^2}{8} \int e_\mu^{\;a} e_\nu^{\;b} \sqrt{-g} R^{\alpha\beta}_{\;\;\;\;ab}  \varepsilon_{\alpha\beta\gamma\delta} \dx{x}^\mu \wedge \dx{x}^\nu \wedge \dx{x}^\gamma \wedge \dx{x}^\delta
        \\
        & = \frac{M_{(4)}^2}{8} \int e_\mu^{\;a} e_\nu^{\;b} \sqrt{-g} R^{\alpha\beta}_{\;\;\;\;ab}  \varepsilon_{\alpha\beta\gamma\delta} \frac{\varepsilon^{\mu\nu\gamma\delta}}{\sqrt{-g}} (\hodge{1})
        \\
        &= \frac{M_{(4)}^2}{8} \int R^{\alpha\beta}_{\;\;\;\;\mu\nu} \left(4\delta^\mu_{[\alpha}\delta^\nu_{\beta]}\right) (\hodge{1})
        \\
        &= \frac{M_{(4)}^2}{2} \int R^{\alpha\beta}_{\;\;\;\;\alpha\beta} (\hodge{1})
        \\
        &= \frac{M_{(4)}^2}{2} \int \dnx{4}{x} \sqrt{-g} R \; ,
    \end{split}
\end{equation*}
where in the \nth{4} line we used the fact that the Riemann tensor is antisymmetric on its last two (2-form) indices.

The action Eq. \eqref{CWGravityAction} is the same as the standard dRGT action for ghost-free multi-metric gravity \cite{dRGT_1,dRGT_2,Cycles,GhostFreedom} in the metric formulation provided that the Deser-van Niewenhuizen symmetric vierbein condition,
\begin{equation}\label{SymmetricVierbeinCondition}
    e_{\;\;\;\;\;a}^{(i)\mu}e_\mu^{(j)b} = e^{(i)\mu b}e^{(j)}_{\mu a} \; ,
\end{equation}
is satisfied. Taking nearest neighbour interactions ensures that this is true, and so avoids the presence of Boulware-Deser ghosts in the clockwork theory \cite{GhostFreedom, Cycles}.

It is important to stress that the multi-gravity theory we are considering is \emph{entirely} specified by a choice for both the number of sites and the potential coefficients $T_{ijkl}$. Therefore, all the results we present in this Section will hold for a general multi-gravity theory with nearest neighbour interactions (and so any multi-gravity devoid of the Boulware-Deser ghost), although we shall later specialise to some choice for the coefficients corresponding to a clockwork model.

\subsection{Modified Einstein Equations}

We get the equations of motion by varying the action with respect to the $i$-th tetrad \cite{Cosmo_spin_2}, $e^{(i)a}$, resulting in:
\begin{equation}\label{CWeom}
    \frac{M_{(4)i}^2}{2} e^{(i)b} \wedge \hodge{R^{(i)}_{ab}} + \varepsilon_{abcd} \sum_{jkl} \mathcal{P}(i) T_{ijkl} e^{(j)b} \wedge e^{(k)c} \wedge e^{(l)c} = \hodge T^{(i)}_a \; .
\end{equation}
$\mathcal{P}(i)$ counts the number of times $(i)$ appears in the coefficient $T_{ijkl}$ i.e. terms of the form
\begin{itemize}
    \itemsep0em
    \item $T_{ijkl} \implies \mathcal{P}(i)=1$
    \item $T_{iijk} \implies \mathcal{P}(i)=2$
    \item $T_{iiij} \implies \mathcal{P}(i)=3$
    \item $T_{iiii} \implies \mathcal{P}(i)=4$ ,
\end{itemize}
and $\hodge T^{(i)}_a$ is the dual of the energy-momentum 1-form $T^{(i)}_a$ associated with matter coupled to the $i$-th site, defined as:
\begin{equation}
    \hodge T^{(i)}_a = \frac{\delta S_M}{\delta e^{(i)a}} = \abs{e^{(i)}} T^{(i)\mu}_{\;\;\;\;\;\;a} \varepsilon_{\mu\nu\alpha\beta}\dx{x}^\nu\wedge\dx{x}^\alpha\wedge\dx{x}^\beta \; ,
\end{equation}
with $T^{(i)\mu}_{\;\;\;\;\;\;\nu} = T^{(i)\mu}_{\;\;\;\;\;\;a} e^{(i)a}_\nu$ and $\abs{e^{(i)}}$ the vierbein determinant. If the matter sector has a metric formulation, this is just the standard energy-momentum tensor of GR.

In components, after applying the Hodge star, Eq. \eqref{CWeom} reads:
\begin{equation}
    M^2_{(4)i} G^{(i)\mu}_{\;\;\;\;\;\nu} + 24\vbein{i}{\nu}{a}\ivbein{i}{\mu}{[a}\ivbein{i}{\lambda_1}{b}\ivbein{i}{\lambda_2}{c}\ivbein{i}{\lambda_3}{d]} \sum_{jkl} \mathcal{P}(i) T_{ijkl} \vbein{j}{\lambda_1}{b}\vbein{k}{\lambda_2}{c}\vbein{l}{\lambda_3}{d} = T^{(i)\mu}_{\;\;\;\;\;\;\nu} \; ,
\end{equation}
and for brevity we can package up the interaction term into a single tensor $W^{(i)\mu}_{\;\;\;\;\;\;\nu}$, so that the Einstein equations are:
\begin{equation}
    M^2_{(4)i} G^{(i)\mu}_{\;\;\;\;\;\nu} + W^{(i)\mu}_{\;\;\;\;\;\;\nu} = T^{(i)\mu}_{\;\;\;\;\;\;\nu} \; ,
\end{equation}
with, explicitly:
\begin{equation}
    W^{(i)\mu}_{\;\;\;\;\;\;\nu} = 24\vbein{i}{\nu}{a}\ivbein{i}{\mu}{[a}\ivbein{i}{\lambda_1}{b}\ivbein{i}{\lambda_2}{c}\ivbein{i}{\lambda_3}{d]} \sum_{jkl} \mathcal{P}(i) T_{ijkl} \vbein{j}{\lambda_1}{b}\vbein{k}{\lambda_2}{c}\vbein{l}{\lambda_3}{d} \; .
\end{equation}

With our nearest neighbour restriction for the interactions, the $W$-tensor expands out as:
\begin{equation}
\begin{split}
    W^{(i)\mu}_{\;\;\;\;\;\;\nu} &= 24\times 4\vbein{i}{\nu}{a}\ivbein{i}{\mu}{[a}\ivbein{i}{\lambda_1}{b}\ivbein{i}{\lambda_2}{c}\ivbein{i}{\lambda_3}{d]}
    \\
    \times &\Bigl[ \Tz \vbein{i}{\lambda_1}{b}\vbein{i}{\lambda_2}{c}\vbein{i}{\lambda_3}{d} 
    \\
    &+3\Ti \vbein{i+1}{\lambda_1}{b}\vbein{i}{\lambda_2}{c}\vbein{i}{\lambda_3}{d}
    \\
    &+3\Tii \vbein{i+1}{\lambda_1}{b}\vbein{i+1}{\lambda_2}{c}\vbein{i}{\lambda_3}{d} 
    \\
    &+\Tiii \vbein{i+1}{\lambda_1}{b}\vbein{i+1}{\lambda_2}{c}\vbein{i+1}{\lambda_3}{d} 
    \\
    &+\tiii \vbein{i-1}{\lambda_1}{b}\vbein{i-1}{\lambda_2}{c}\vbein{i-1}{\lambda_3}{d}
    \\
    &+3\tii \vbein{i-1}{\lambda_1}{b}\vbein{i-1}{\lambda_2}{c}\vbein{i}{\lambda_3}{d}
    \\
    &+3\ti \vbein{i-1}{\lambda_1}{b}\vbein{i}{\lambda_2}{c}\vbein{i}{\lambda_3}{d} \Bigr]
    \; .
\end{split}
\end{equation}

\subsection{Constraints and Energy Conservation}

Because of the individual diffeomorphisms associated with each site, each Einstein tensor is covariantly conserved with respect to its own (Levi-Civita) connection:
\begin{equation}\label{Bianchi}
    \nabla^{(i)\mu}G^{(i)}_{\mu\nu} = 0 \; ,
\end{equation}
but because of our new interaction terms, this means that, individually, the energy-momentum tensors are not covariantly conserved, but satisfy instead:
\begin{equation}\label{T_W_relation}
    \nabla^{(i)\mu}T^{(i)}_{\mu\nu} = \nabla^{(i)\mu}W^{(i)}_{\mu\nu} \; .
\end{equation}
However, because we also have our overall diagonal diffeomorphism invariance (which gives rise to the clockwork zero-mode), the matter sector as a whole, over all sites, is conserved:
\begin{equation}\label{Matter conservation}
    \sum_{i=0}^{N-1} \abs{e^{(i)}}\nabla^{(i)\mu}T^{(i)}_{\mu\nu} = 0 \; ,
\end{equation}
which therefore implies the following constraint on the interactions due to Eq. \eqref{T_W_relation}, which we shall henceforth refer to as the \emph{Bianchi constraint}:
\begin{equation}\label{W constraint}
    \sum_{i=0}^{N-1} \abs{e^{(i)}}\nabla^{(i)\mu}W^{(i)}_{\mu\nu} = 0 \; .
\end{equation}

The Bianchi constraint tells us something about what is going on physically. For example, if we have matter coupled to only one site, then by virtue of Eq. \eqref{Matter conservation} the energy-momentum tensor on that site \emph{is} conserved individually and so a free test particle would follow the geodesics of that site's metric. By Eq. \eqref{T_W_relation}, the corresponding $W$-tensor, and as a consequence every other $W$-tensor, is also conserved individually in this case. The force between two sources, however, will comprise a contribution from the zero-mode, as well as suppressed contributions from the additional massive modes. If matter couples to more than one site, it is not obvious what the `physical' metric should be and there is more work to be done to try and understand this.

Regardless, we can now begin to use all this technology to try to solve our modified Einstein equations for cosmological FRW-like solutions.

\subsection{Vacuum solutions}\label{Sec:Static_discrete}

The first situation we consider is the vacuum case, where there is no matter on any of the sites i.e. $S_M=0$. We shall take the ansatz for the gear metrics to be:
\begin{equation}\label{dyn_metric}
    \dx{s}^2_{(i)} = -c_i^2(t) \dx{t}^2 + a_i^2(t) \eta_{jk}\dx{x}^j\dx{x}^k \; ,
\end{equation}
and eventually look for deSitter vacuum solutions of the form $a_i(t) = a_{i,0} e^{H_i t}$ for some set of constant Hubble parameters $H_i$, and $a_{i,0}=a_i(0)$. Since all of the metrics live on the same space, we can only rescale the coordinates to fix the lapse and scale factor of \emph{one} of the metrics, while the rest must remain free (normally we will choose to fix them to 1 on an appropriate site e.g. the first site, in vacuum, or the site where matter couples, if there is a matter coupling).

With this choice for our metric, the Einstein tensor has the following non-vanishing components:
\begin{align}
    G^{(i)0}_{\;\;\;\;\;0} &= -3\left(\frac{\dot{a}_i}{a_i c_i}\right)^2
    \\
    G^{(i)j}_{\;\;\;\;\;k} &= \frac{1}{c_i^2}\left(-\frac{\dot{a}^2_i}{a^2_i} - 2\frac{\ddot{a}_i}{a_i} + 2\frac{\dot{a}_i}{a_i}\frac{\dot{c}_i}{c_i}\right) \delta^j_k \; .
\end{align}
Also, the vierbeins are:
\begin{align}
    \vbein{i}{0}{0} &= c_i
    \\
    \vbein{i}{j}{k} &= a_i\delta^k_j \; ,
\end{align}
which lead to the following non-vanishing $W$-tensor components, in terms of the $\alpha$'s and $\beta$'s defined in Section \ref{Sec:Formalism} for the potential coefficients:
\begin{equation}\label{W00}
    W^{(i)0}_{\;\;\;\;\;\;0} = \alpha_i \sum_{m=0}^{4} 24\beta_m \binom{3}{m}a_{i+1}^m a_i^{-m} + \alpha_{i-1} \left(\frac{a_{i-1}}{a_i}\right)^4  \sum_{m=0}^{4} 24\beta_m \binom{3}{m-1} a_{i-1}^{-m}a_i^m
\end{equation}
\begin{equation}\label{Wij}
    \begin{split}
        W^{(i)j}_{\;\;\;\;\;\;k} = 24\delta^j_k \Bigl\{ \alpha_i &\bigl[  \beta_0 + \beta_1 (c_{i+1}c_i^{-1} + 2a_{i+1}a_i^{-1}) + \beta_2(2c_{i+1}c_i^{-1}a_{i+1}a_i^{-1} + a_{i+1}^2a_i^{-2})
        \\
        &+ \beta_3(c_{i+1}c_i^{-1}a_{i+1}^2a_i^{-2})\bigr]
        \\
        &+ \alpha_{i-1}\bigl[ \beta_1(c_{i-1}c_i^{-1}a_{i-1}^2a_i^{-2}) + \beta_2 (2c_{i-1}c_i^{-1}a_{i-1}a_i^{-1} + a_{i-1}^2a_i^{-2})
        \\
        &+ \beta_3 (c_{i-1}c_i^{-1} + 2a_{i-1}a_i^{-1}) + \beta_4 \bigr] \Big\} \; .
    \end{split}
\end{equation}

So, we have 2 sets of Einstein equations: a modified Friedmann equation coming from all of the 00-type terms, and a modified Raychaudhuri equation coming from the $jk$-type terms.

Considering first the modified Friedmann equation, and substituting in our deSitter ansatz for the vacuum solution, we get:
\begin{equation}\label{Friedmann_vac}
\begin{split}
    3M_{(4)i}^2 \left(\frac{H_i}{c_i}\right)^2 = &\alpha_i \sum_{m=0}^{4} 24\beta_m \binom{3}{m}a_{i+1,0}^m a_{i,0}^{-m}e^{m(H_{i+1}-H_i)t} \\&+ \alpha_{i-1} \sum_{m=0}^{4} 24\beta_m \binom{3}{m-1} a_{i-1,0}^{4-m}a_{i,0}^{m-4} e^{(4-m)(H_{i-1}-H_i)t} \; .
\end{split}
\end{equation}
Note that we still we have the lapse on the left hand side that we must deal with. Thankfully, we are able to make some progress here due to the Bianchi constraint, Eq. \eqref{W constraint}. In Appendix \ref{App:Bianchi constraint}, we show that if there is no matter coupling, or matter couples to one site only, then the only way to satisfy the constraint is to take the lapses to be given in terms of the scale factors as,
\begin{equation}\label{lapses_discrete}
     c_i = \frac{\dot{a}_i}{\dot{a}_I} \; ,
\end{equation}
where $I$ is the site whose lapse we fix to 1 via coordinate rescaling. In terms of Eq. \eqref{W constraint} this is the case where each term in the sum vanishes separately i.e. where every $W$-tensor is covariantly conserved with respect to its own connection, which must necessarily be the case in vacuum. This directly generalises the solution for 2 sites given in \cite{Cosmo_spin_2,Bimetric} and for 3 sites in \cite{Cosmo_3_spin2} (who work in conformal time so that the above becomes $c_i=(\dot{a}_i/\dot{a}_I)a_I$, but it describes the same situation). Physically, it means that there is no flow of energy-momentum across the sites. The solution also automatically satisfies the Bianchi constraint \emph{even if} matter couples to more than one site; it is just that more complicated solutions could also exist, in that scenario.

With this expression for the lapses, something nice happens to the Friedmann equation. Substituting into Eq. \eqref{Friedmann_vac} yields:
\begin{equation}\label{Friedmann_no_lapse}
\begin{split}
    3M_{(4)i}^2 H_0^2 = a_{i,0}^2 e^{2(H_i-H_0)t} \bigg[ &\alpha_i \sum_{m=0}^{4} 24\beta_m \binom{3}{m}a_{i+1,0}^m a_{i,0}^{-m}e^{m(H_{i+1}-H_i)t} \\ &  + \alpha_{i-1} \sum_{m=0}^{4} 24\beta_m \binom{3}{m-1} a_{i-1,0}^{4-m}a_{i,0}^{m-4} e^{(4-m)(H_{i-1}-H_i)t}\bigg] \; ,
\end{split}
\end{equation}
and we see that the LHS is now a constant, while the RHS is time-dependent. The only way that these equations can be satisfied is if all sites have the same Hubble parameter, thus killing the time dependence by forcing the exponentials on the RHS to 1! We find, therefore, that the theory possesses deSitter vacua,
\begin{equation}
    a_i(t) = a_{i,0} e^{Ht} \; ,
\end{equation}
for some constant $H$ (so the lapses are, explicitly, $c_i=a_{i,0}/a_{I,0}$ -- in particular this means that the Friedmann and Raychaudhuri equations become equivalent), where the $a_{i,0}$'s and $H$ satisfy the \emph{algebraic} equations:
\begin{equation}\label{dS vacua}
    3M_{(4)i}^2 H^2 = a_{i,0}^2 \bigg[ \alpha_i \sum_{m=0}^{4} 24\beta_m \binom{3}{m}a_{i+1,0}^m a_{i,0}^{-m} \\   + \alpha_{i-1} \sum_{m=0}^{4} 24\beta_m \binom{3}{m-1} a_{i-1,0}^{4-m}a_{i,0}^{m-4}\bigg] \; .
\end{equation}
Since we are free to set $a_{I,0}=1$, this is a system of $N$ equations for $N$ variables ($H$ and the $N-1$ remaining $a_{i,0}$'s) and is hence in principle solvable, with a number of solutions, corresponding to deSitter vacua with different values of $H$. The number of physical deSitter vacua depends only on the number of solutions to these equations that have real scale factors, which in general is dependent on both the number of sites and the potential coefficients $T_{ijkl}$.

Naturally, the vacuum condition Eq. \eqref{FullVacCond} we derived in Section \ref{Sec:Formalism} is simply a special case of Eq. \eqref{dS vacua}\footnote{It is important to stress, however, that without the symmetric polynomial formalism we would not have possessed the intuition about splitting the $T_{ijkl}$ into $\alpha$'s and $\beta$'s that was crucial in all of this.}. Namely, it is the static (i.e. $H=0$) solution where we impose the desired clockwork vacuum structure $a_{i,0} = a_{0,0}/q^i$ (if we set $a_{I,0}=1$ then this also fixes the overall normalisation to be $a_{0,0}=q^I$), and then choose our $\alpha$'s and $\beta$'s to ensure that this vacuum is indeed a solution to the Einstein equations\footnote{A choice of $\alpha$'s and $\beta$'s completely specifies the theory. We choose them such that the clockwork vacuum is a static solution to our theory, but in principle one could choose them however they wish -- the results of this section are entirely general.}. In fact, if we work in conformal time (so that the lapses become $c_i = a_{i,0}$, and all metrics are conformally flat) one can see directly that the clockwork gravity potential Eq. \eqref{CWGPotential} is identical to the $\deg=4$ scalar potential Eq. \eqref{Potential} we used throughout Section \ref{Sec:Formalism}, with the conformal factors $a_{i,0}$ playing the role of the scalars $\phi_i$. 

The upshot is that we are free to use all of the techniques we have already developed in order to pick a good set of coefficients $T_{ijkl}$ for the gravitational theory. This carries over to the graviton mass matrix too -- which we can determine by expanding around the static vacuum solution,
\begin{equation}\label{expanded metric}
    g^{(i)}_{\mu\nu} = a_{i,0}^2\eta_{\mu\nu} + \frac{a_{i,0}}{M_{(4)i}}h^{(i)}_{\mu\nu} \; ,
\end{equation}
where the normalisation by $M_{(4)i}$ is to ensure the Fierz-Pauli kinetic term for $h^{(i)}_{\mu\nu}$ is canonical \cite{FierzPauli}. The second-order variation of the potential is then \cite{ClockworkGrav}:
\begin{equation}
    S^{(2)}_V = \frac12\int \dnx{4}{x} \sum^{N-1}_{i,j=0}\frac{1}{2M_{(4)i}M_{(4)j}} M^2_{ij} \left[ h^{(i)}h^{(j)} - h^{(i)\mu}_{\;\;\;\;\;\nu}h^{(i)\nu}_{\;\;\;\;\;\mu}\right] \; ,
\end{equation}
where $h^{(i)}=h^{(i)\mu}_{\;\;\;\;\;\mu}$, and the mass matrix $M^2_{ij}$ is as in Section \ref{Sec:Formalism}, so we are able to calculate its components given a set of $T_{ijkl}$, and hence determine the spectrum of graviton masses.

\subsubsection{The physical Planck scale}

We have thus far not said anything about how we should interpret the quantity $M_{(4)i}$, which looks like a Planck scale for each site, so might naively be assumed to be just that. The physical Planck scale $M_{\text{eff}}$, however, is the one associated with the clockwork zero-mode, which is related to $M_{(4)i}$ but is crucially \emph{not the same thing}. 

To figure out what this scale should be, we must work in terms of the graviton mass eigenstates, which we obtain as in Section \ref{Sec:QCS} via an orthogonal rotation of the field basis,
\begin{equation}
    h^{(i)}_{\mu\nu} = \sum_{j=0}^{N-1} O^{ij} \tilde{h}^{(j)}_{\mu\nu} \; ,
\end{equation}
where the columns of the orthogonal matrix $O$ are given by the mass eigenvectors. In particular, we have the zero-mode, $\tilde h_{\mu\nu}^{(0)}=h^{(i)}_{\mu\nu}O^{i0} = \frac{\mathcal{N}}{q^i}h^{(i)}_{\mu\nu}$, where the normalisation is:
\begin{equation}
    \mathcal{N} = \frac{1}{\sqrt{\sum_{i=0}^{N-1} q^{-2i}}} = \sqrt{\frac{1-q^{-2}}{1-q^{-2N}}} \; .
\end{equation}
Note that for $q>1$ we see that the contribution to the zero-mode from the $i$-th metric diminishes as $i$ increases towards $N-1$.
It also follows that the gear metrics may be written as:
\begin{equation}
    g^{(i)}_{\mu\nu} = a_{i,0}^2\eta_{\mu\nu} + \frac{a_{i,0}}{M_{(4)i}}\frac{\mathcal{N}}{q^i} \tilde h^{(0)}_{\mu\nu} + \frac{a_{i,0}}{M_{(4)i}} \sum_{j=1}^{N-1} O^{ij} \tilde h^{(j)}_{\mu\nu} \; .
\end{equation}

So, if we include a minimal coupling to matter on the $I$-th site, and assume (as in \cite{ClockworkGrav}) that the fundamental clockwork scale $M_{(4)i}=M_{(4)}$ is the same on all sites, then the variation in the action becomes (fixing $a_{I,0}=1$ so that $a_{0,0}=q^I$):
\begin{equation}\label{massive_modes}
\begin{split}
    \delta S_M &= \frac12 \int \dnx{4}{x} \; \delta g^{(I)}_{\mu\nu} T^{(I)\mu\nu}
    \\
    &=  \int \dnx{4}{x} \; \frac{1}{2M_{\text{eff}}} \tilde h^{(0)}_{\mu\nu} T^{(I)\mu\nu} + \frac{1}{2M_{(4)}}\sum_{j=1}^{N-1} O^{ij} \tilde h^{(j)}_{\mu\nu} \; ,
\end{split}
\end{equation}
and we can identify the physical Planck scale as:
\begin{equation}\label{zero_mode_coupling}
    M^2_{\text{eff}} = \frac{1-q^{-2N}}{1-q^{-2}}q^{2I} M_{(4)}^2 \; ,
\end{equation}
which can be much larger than $M_{(4)}$ if the number of fields in the chain is big enough (indeed, this is the purpose of using the clockwork).

\subsubsection{Adding a cosmological constant}

The full treatment of the Einstein equations given above is more general than the derivation of our vacuum condition Eq. \eqref{FullVacCond} from Section \ref{Sec:Formalism}, and allows for modification by a cosmological constant. We see this by taking on some sites $j$ a non-zero energy-momentum tensor of the form:
\begin{equation}
    T^{(j)}_{\mu\nu} = -\sigma_j g^{(j)}_{\mu\nu} \; ,
\end{equation}
for some constants $\sigma_j$, which will subsequently appear on the RHS of our Einstein equations. In particular, the static $H=0$ clockwork vacuum solution which we use as a means of choosing a good set of potential coefficients $T_{ijkl}$ is modified to:
\begin{equation}\label{vacuum_condition_sigmas}
    \alpha_i \sum_{m=0}^{\deg} \deg!\beta_m \binom{\deg-1}{m}q^{-m} + q^{\deg} \alpha_{i-1}  \sum_{m=0}^{\deg} \deg!\beta_m \binom{\deg-1}{m-1}q^{-m} + \sum_j \sigma_j \delta^j_i = 0 \;\;\; \forall i \;,
\end{equation}
where we write the result for general $\deg$, but of course we have $\deg=4$.

A particularly interesting case is the one where we have $j=0, N-1$ i.e. we place a cosmological constant on only the first and last sites of the clockwork lattice. In this case, since we have (by definition) that $\alpha_{-1}=\alpha_{N-1}=0$, the two $\sigma$'s play the role of the missing sum on each of the end sites, that is:
\begin{align}
    \sigma_0 &=  -\alpha_0 \sum_{m=0}^{\deg} \deg!\beta_m \binom{\deg-1}{m}q^{-m} \; ,
    \\
    \sigma_{N-1} &=  -q^{\deg}\alpha_{N-2} \sum_{m=0}^{\deg} \deg!\beta_m \binom{\deg-1}{m-1}q^{-m} \; .
\end{align}
If all of the nonzero $\alpha$'s are equal, the vanishing of the bulk equations implies that the $\sigma$'s must be equal and opposite:
\begin{equation}
    \alpha_n = \alpha \; \forall n \implies \sigma_0 = - \sigma_{N-1} \; .
\end{equation}
In Section \ref{Sec:Continuum}, we will see that in the continuum limit these $\sigma$'s are identified with the respective tensions of branes placed either end of the clockwork lattice, so this result is not surprising.

\subsection{Matter solutions}

Now we wish to add a minimal matter coupling to some of the sites. Since we are interested in cosmology, we assume that each energy-momentum tensor is of perfect fluid form, but keep in the possibility of an additional cosmological constant $\sigma$, i.e.
\begin{equation}\label{perfect_fluid_Tmunu}
    T^{(i)}_{\mu\nu} = (\rho_i + \sigma_i)u_\mu u_\nu + (p_i-\sigma_i)\gamma^{(i)}_{\mu\nu} \; ,
\end{equation}
for 4-velocity $u_\mu$ and $\gamma^{(i)}_{\mu\nu} = u_\mu u_\nu + g^{(i)}_{\mu\nu}$.

When $T^{(i)}_{\mu\nu}$ refers to any form of matter other than a cosmological constant (e.g. radiation, pressureless dust), it is necessarily time-dependent due to the conservation equation \eqref{Matter conservation}. This means that our exponential ansatz for the vacuum solutions no longer works, because the RHS of our Einstein equations will \emph{always} be time-dependent, which is inconsistent with the constancy of the LHS. Therefore, we must work with the equations in general, and figure out a way to solve them.

Written out in full, the modified Friedmann equation reads:
\begin{equation}\label{Friedmann}
\begin{split}
    3M_{(4)i}^2 \left(\frac{\dot{a}_i}{a_i c_i}\right)^2 = &\alpha_i \sum_{m=0}^{4} 24\beta_m \binom{3}{m}a_{i+1}^m a_i^{-m} \\&+ \alpha_{i-1} \left(\frac{a_{i-1}}{a_i}\right)^4  \sum_{m=0}^{4} 24\beta_m \binom{3}{m-1} a_{i-1}^{-m}a_i^m  + (\rho_i + \sigma_i) \; ,
\end{split}
\end{equation}
and the modified Raychaudhuri equation is:
\begin{equation}\label{Raychaudhuri}
\begin{split}
    \frac{M_{(4)i}^2}{c_i^2} \left(\frac{\dot{a}^2_i}{a^2_i} + 2\frac{\ddot{a}_i}{a_i} - 2\frac{\dot{a}_i}{a_i}\frac{\dot{c}_i}{c_i}\right) \\= 24 \Bigl\{ \alpha_i &\bigl[  \beta_0 + \beta_1 (c_{i+1}c_i^{-1} + 2a_{i+1}a_i^{-1}) + \beta_2(2c_{i+1}c_i^{-1}a_{i+1}a_i^{-1} + a_{i+1}^2a_i^{-2})
        \\
        &+ \beta_3(c_{i+1}c_i^{-1}a_{i+1}^2a_i^{-2})\bigr]
        \\
        &+ \alpha_{i-1}\bigl[ \beta_1(c_{i-1}c_i^{-1}a_{i-1}^2a_i^{-2}) + \beta_2 (2c_{i-1}c_i^{-1}a_{i-1}a_i^{-1} + a_{i-1}^2a_i^{-2})
        \\
        &+ \beta_3 (c_{i-1}c_i^{-1} + 2a_{i-1}a_i^{-1}) + \beta_4 \bigr] \Big\} - (p_i - \sigma_i) \; .
\end{split}
\end{equation}

At first glance, these equations appear very ugly: they are $N$ highly nonlinear coupled differential equations, which are difficult to solve (and potentially impossible to do so analytically in general). To make things more tractable, we can use the same result for the lapses that we used in deriving the vacuum solutions, since this came directly from the Bianchi constraint, which still holds here. Therefore, we fix the lapse/scale factor to 1 on some site $I$ where there is a minimal matter coupling and so take $c_i = \dot{a}_i/\dot{a}_I$.

Substituting these lapses into the Friedmann equations, Eqs. \eqref{Friedmann}, as before, yields a series of equations for $\dot{a}_I$, which we package together as:
\begin{equation}\label{adot0}
    3M_{(4)i}^2\left(\frac{\dot{a}_I}{a_i}\right)^2 = f_i(\mathbf{a}) \; ,
\end{equation}
where each $f_i$ is understood as the RHS of the $i$-th Friedmann equation. That is, $f_i$ essentially represents the 00-component of the $i$-th $W$-tensor encoding the clockwork interactions, plus any additional matter minimally coupled to the $i$-th site. Each $f_i$ is a function of $a_i$ and its nearest-neighbours only i.e. $f_i(\mathbf{a})=f_i(a_i,a_{i-1},a_{i+1})$. While we only have an evolution equation for $a_I$, taking the ratio of $f_i$ and $f_j$ gives a set of algebraic conditions that the $a$'s must satisfy throughout the evolution, namely, that:
\begin{equation}\label{fi_fj}
    a_j^2 f_j(\mathbf{a}) = a_i^2 f_i(\mathbf{a})\; .
\end{equation}
So, we should (at least numerically) be able to evolve $a_I$ via Eq. \eqref{adot0}, and at each time step ensure that the rest of the scale factors obey Eq. \eqref{fi_fj}. This will then implicitly track the evolution of all $a_i$, and so solve for the background evolution. We will do this to solve the 
 evolution equations for two example models in Section \ref{Sec:Examples}. 
 
 As eluded to in Section \ref{Sec:Intro}, one of the models we use as a consistency check for our work is a deconstructed version of the RS1 braneworld. It is not immediately obvious that we can do this, so it is useful to first review how we can relate our discrete clockwork to a continuum theory in 5D, of which RS1 exists as a special case, with suitable modifications to the system boundaries. This is explored in the next Section, clarifying the work of \cite{Deconstructing}.

\section{Continuum Clockwork Gravity}\label{Sec:Continuum}

Following \cite{Deconstructing}, we can relate our 4D clockwork quantities to an underlying 5D geometry into which they are embedded. We take the continuum 5D line element to be:
\begin{equation}
    \dx{s}^2 = g_{\mu\nu}(x,y) + \dx{y}^2 \; ,
\end{equation}
which corresponds to a 5D geometry given by $\mathcal{M} = \mathbf{M}_4 \times [0,L]$, where $\mathbf{M}_4$ is our usual 4D spacetime, parametrised by coordinates $x^\mu$, and the new coordinate $y\in [0,L]$ parametrises the compact extra dimension, which lives on an interval from 0 to $L$. Our bulk spacetime $\mathcal{M}$ possesses a boundary $\partial\mathcal{M}$ that has two components: one at $y=0$ and one at $y=L$. The component at $y=0$ is negatively oriented, whereas the component at $y=L$ is postively oriented, in the sense that integrating over the boundary component in question comes equipped with the appropriate sign. Usually, when one talks of these kind of 5D deconstructions, the extra dimension is orbifolded on $S_1/\mathbb{Z}_2$ \cite{RS1,RS2,Carsten_review,Langlois_review}, with fixed points of $\mathbb{Z}_2$ at $y=0$ and $y=L$ rather than true boundaries, with the $\mathbb{Z}_2$-symmetry being used to ease calculations at these special points. With the clockwork, we are not afforded this luxury. The reason for this will become clear very shortly, but for now we work with the geometry as described. From now on we will use $M,N=0,1,2,3,5$ to refer to 5D indices and $\mu,\nu=0,1,2,3$ to refer to 4D indices.

With our choice of line element, 4D hypersurfaces of constant $y$ have a very simple unit normal, $n^M=(\mathbf{0}, 1)$, and so the induced metric $h_{MN}=g_{MN}-n_M n_N$ has only $h_{\mu\nu}=g_{\mu\nu}$ as its non-zero components. We can define the constant-$y$ hypersurfaces' extrinsic curvature for arbitrary vector fields $X$ and $Y$ by:
\begin{equation}
    K(X,Y) = g(\nabla_X n, Y) \; .
\end{equation}
The components are given by the Lie derivative of $h_{MN}$ along the normal vector, $K_{MN}=\frac12 \mathcal{L}_n h_{MN}$, so only the following components are non-vanishing:
\begin{equation}\label{ExtrinsicCurvature}
    K_{\mu\nu} = \frac12 \partial_y g_{\mu\nu} \; ,
\end{equation}
which in terms of the vierbeins reads:
\begin{equation}\label{ExtrinsicCurvatureVierbeins}
    K^\mu_{\;\;\nu}= \frac12 e^{\mu b} e^\sigma_{\;\;b} e_{\nu a} \partial_y e_\sigma^{\;\;a} + \frac12 e^{\mu a} \partial_y e_{\nu a} \; .
\end{equation}

We connect the discrete theory to the continuum by interpreting the $(i)$ indices as corresponding to discrete locations in the \nth{5} dimension, separated by some distance $\delta y$. That is, we have $y_i=i\delta y$, and the $i$-th discrete clockwork metric is the induced metric on the hypersurface at $y_i$:
\begin{equation}
    g^{(i)}_{\mu\nu}(x) = g_{\mu\nu}(x, y_i) \; .
\end{equation}
This picture makes it clear why we cannot orbifold our \nth{5} dimension on $S_1/\mathbb{Z}_2$: the first and last sites of the clockwork lattice in the discrete theory are just that, endpoints of the system, so in the continuum limit they must become true boundaries of the extra dimension, without $\mathbb{Z}_2$-symmetry. When we come to talk about RS1, we shall see the effect of this lack of $\mathbb{Z}_2$-symmetry on calculations explicitly.

We are lead to introduce finite difference expressions for derivatives in the $y$ direction,
\begin{equation}
    \partial_y e_{\mu a} \rightarrow \frac{1}{\delta y} \left[ e^{(i+1)}_{\mu a} - e^{(i)}_{\mu a} \right] \; ,
\end{equation}
which we can then use, along with the symmetric vierbein condition (and a lot of algebra), to rewrite the discrete action Eq. \eqref{CWGravityAction} in terms of these extrinsic curvatures. The continuum limit is achieved by sending $\delta y \rightarrow 0$ and the number of sites $N\rightarrow\infty$, while keeping the product $(N-1)\delta y=L$ fixed. In this limit, the clockwork action 
 Eq. \eqref{CWGravityAction} becomes:
\begin{equation}\label{CWGravityContinuum}
    S_{\text{bulk}} = \int_\mathcal{M} \left[ \frac{M^3_{(5)}}{2} R_{(5)} - 2\Lambda_5(y) + \alpha_1(y)M^4_{(5)}K  + \alpha_2(y)M^3_{(5)}K_2  + \alpha_3(y)M^2_{(5)}K_3  \right] \; ,
\end{equation}
where
\begin{align}
    K &= K^\mu_{\;\;\mu}
    \\
    K_2 &= \delta^{[\mu}_\alpha \delta^{\nu]}_\beta K^\alpha_{\;\;\mu} K^\beta_{\;\;\nu}
    \\
    K_3 &= \delta^{[\mu}_\alpha \delta^{\nu}_\beta \delta^{\rho]}_\gamma K^\alpha_{\;\;\mu} K^\beta_{\;\;\nu} K^\gamma_{\;\;\rho}
    \\
    2\Lambda_5 (y) &= \frac{24}{\delta y} \left( \Tz + 4\Ti + 6\Tii + 4 \Tiii \right)
    \\
    \alpha_1 (y) M^4_{(5)} &= -24 \left( \Ti + 3\Tii -3\Tiii \right)
    \\
    \alpha_2 (y) M^3_{(5)} + M^3_{(5)} &= -24\delta y \left( \Tii + 2\Tiii \right)
    \\   
    \alpha_3 (y) M^2_{(5)} &= -24\delta y^2 \Tiii
    \\
    M^3_{(5)} &= \frac{M^2_{(4)}}{\delta y} \; ,
\end{align}
and $R_{(5)}$ is the Ricci scalar constructed from the 5D metric with components $g_{MN}$. If one wishes, the extrinsic curvature terms can instead be realised as an extra scalar degree of freedom, which is not a dilaton \cite{Deconstructing}. We can also invert these to give us the coefficients $T_{ijkl}$ in terms of the 5D quantities\footnote{We corrected a minus sign error in \cite{Deconstructing} here.}:
\begin{align}
    24\Tz &= 2\Lambda_5 \delta y + 28\alpha_3 \frac{M^2_{(5)}}{\delta y^2} - 6 \frac{\alpha_2 M^3_{(5)} + M^3_{(5)}}{\delta y} + 4\alpha_1 M^4_{(5)}\label{Tz}
    \\
    24\Ti &= -9\alpha_3 \frac{M^2_{(5)}}{\delta y^2} +3 \frac{\alpha_2 M^3_{(5)} + M^3_{(5)}}{\delta y} - \alpha_1 M^4_{(5)}\label{Ti}
    \\
    24\Tii &= 2\alpha_3 \frac{M^2_{(5)}}{\delta y^2} - \frac{\alpha_2 M^3_{(5)} + M^3_{(5)}}{\delta y}\label{Tii}
    \\
    24\Tiii &= -\alpha_3 \frac{M^2_{(5)}}{\delta y^2}\label{Tiii} \; .
\end{align}

Eq. \eqref{CWGravityContinuum} is the action that describes the 5D bulk in the continuum theory. To account for the boundary $\partial\mathcal{M}$, we must include a Gibbons-Hawking term \cite{York,GH} to ensure that the variational problem is well-posed:
\begin{equation}
    S_{\text{boundary}} = \int_{\partial\mathcal{M}} \Mfive K \; ,
\end{equation}
where $K$ is the trace of the extrinsic curvature on the boundary $\partial\mathcal{M}$.

In principle, we may also place a brane at each of the respective boundary components, each with some tension $\sigma_i$ and matter Lagrangian $\mathcal{L}_{m,i}$:
\begin{equation}
    S_{\text{branes}} = \sum_{i=L,R} \int \text{d}^4 x \, \sqrt{-h^{(i)}} \, \left( -\sigma_i + \mathcal{L}_{m,i} \right) \; ,
\end{equation}
where $h^{(i)}$ are the induced metrics on each of the left and right branes -- which we identify with the first and last site of clockwork i.e. $h^{(L)}_{\mu\nu}(x) = g_{\mu\nu}(x, 0) = g_{\mu\nu}^{(0)}$, and $h^{(R)}_{\mu\nu}(x) = g_{\mu\nu}(x,L) = g_{\mu\nu}^{(N-1)}$. The full action for the continuum clockwork is then just the sum of the three pieces outlined above,
\begin{equation}\label{FullContinuumAction}
    S = S_{\text{bulk}} + S_{\text{boundary}} + S_{\text{branes}} \; .
\end{equation}

\subsection{Randall-Sundrum (Continuum)}\label{Sec:RS_Continuum}

The Randall-Sundrum model is the simplest possible special case of the continuum theory \eqref{FullContinuumAction}, which has $\alpha_1=\alpha_2=\alpha_3=0$ and $\Lambda_5=\text{const}$, so that the bulk is just pure 5D gravity i.e.
\begin{equation}
    S_{\text{bulk}} = \int_{\mathcal{M}} \frac{\Mfive}{2} R_{(5)} - 2\Lambda_5 \; .
\end{equation}

To determine the equations of motion, we vary the action with respect to the metrics in question. In standard GR, variations of the boundary metric $h$ vanish, but here we must allow them to be arbitrary, as we are interested in what they do (they correspond to the first and last sites of the clockwork). The result is that we get two sets of equations -- one set for the bulk due to the Ricci scalar variation:
\begin{equation}\label{Bulk}
    \Mfive G_{MN}+2\Lambda_5 g_{MN} = 0 \; ,
\end{equation}
which is just standard GR in 5D, and one set for the boundary due to the Gibbons-Hawking variation, which reads:
\begin{equation}\label{Boundary}
\begin{split}
    K_{MN}-Kh_{MN} &= -\kappa^2 S_{MN} \;\;\;\;\; (y=0)
    \\
    K_{MN}-Kh_{MN} &= +\kappa^2 S_{MN} \;\;\;\;\; (y=L) \; ,
\end{split}
\end{equation}
where $S_{MN}$ is the brane energy-momentum tensor associated with $\mathcal{L}_{m,i}$ and $\sigma_i$, $\kappa^2=1/\Mfive$, and the sign change on the RHS is due to the change in orientation between the two boundary components. 

We can understand these boundary equations as being \emph{one side} of the usual Israel junction conditions across a singular hypersurface embedded in an underlying spacetime manifold \cite{Israel, Shtanov}. Indeed, we can reconstruct the usual $S_1/\mathbb{Z}_2$ orbifold by thinking of the full spacetime in that case to be constructed from two separate intervals, $[-L,0]$ and $[0,L]$, which share a common boundary at $y=0$, and also at $y=L$ due to the $\mathbb{Z}_2$ symmetry. Variation of the Gibbons-Hawking term with this construction leads to precisely the Israel conditions across the branes at $y=0$ and $y=L$, and the $\mathbb{Z}_2$ symmetry manifests as an additional factor 1/2 on the RHS of Eqs. \eqref{Boundary}. For us, we \emph{only} have the interval $[0,L]$, so there is no notion of jumping across a brane, and hence our boundary equations have no factor 1/2.

Since the bulk is just 5D GR, the Bianchi identity says that the Einstein tensor is covariantly conserved. By Eq. \eqref{Bulk}, so too is the energy-momentum tensor:
\begin{equation}
    \nabla^M G_{MN} = \nabla^M T_{MN} = 0 \; .
\end{equation}
As a result of the Codazzi equation, which relates the 5D quantities to their projection onto a 4D hypersurface \cite{Wald,Langlois_review}, the brane energy-momentum tensor is also covariantly conserved with respect to its associated covariant derivative i.e.
\begin{equation}
    \nabla_{(4)}^\mu S_{\mu\nu} = 0 \; .
\end{equation}

The solutions to this system are well-studied in the literature (see e.g. \cite{Langlois_review,Carsten_review,Full_EFEs_brane} and refs therein), so we shall only give a very brief overview.

\subsubsection{Static vacuum solution}

This is the solution where we have no matter, only tension, on the brane i.e. $S_{MN}=-\sigma h_{MN}$, which is the continuum version of the situation in the discete theory where we have only a cosmological constant on each of the end sites. In full analogy with the discrete case, we take our metric to depend only on $y$, and make it such that the hypersurfaces are conformally flat. That is, the bulk is $\text{AdS}_5$, and we have:
\begin{equation}\label{RS1_static}
    \dx{s}^2 = e^{-2A(y)}\eta_{\mu\nu} \dx{x}^\mu\dx{x}^\nu + \dx{y}^2 \; ,
\end{equation}
where for convenience we work with the \emph{warp factor} $A(y)$ rather than the usual scale factor $a(y)$, but of course we can go between them with the identification $a=e^{-A}$. With this metric, the bulk equations become:
\begin{align}
    G_{\mu\nu}&=(6A^{\prime2}-3A^{\prime\prime})g_{\mu\nu}
    \\
    G_{\mu5}&=G_{5\mu}=0
    \\
    G_{55}&=6A^{\prime2} \; ,
\end{align}
where primes denote derivatives with respect to $y$. From the 55-equation we get\footnote{We could, of course, also have had $A=-ky$. All this amounts to is changing perspective on which brane you are looking from.}:
\begin{equation}
    A = ky \; , \;\;\; k^2 = -\frac{\Lambda_5}{3\Mfive} \; ,
\end{equation}
and the $\mu\nu$-equation is internally consistent with this result.

Regarding the boundary, the only surviving components of the extrinsic curvature are $K_{\mu\nu}=-A^\prime h_{\mu\nu}$, so we get:
\begin{equation}
\begin{split}
    3A^\prime h_{\mu\nu} &= +\kappa^2\sigma_0 h_{\mu\nu} \;\;\;\;\; (y=0)
    \\
    3A^\prime h_{\mu\nu} &= -\kappa^2 \sigma_L h_{\mu\nu} \;\;\;\;\; (y=L) \; ,
\end{split}
\end{equation}
which implies that our brane tensions must satisfy:
\begin{equation}
    \sigma_0 = -\sigma_L = 3k\Mfive \; .
\end{equation}
This is the standard RS1 solution, adapted to our lack of $\mathbb{Z}_2$-symmetry (including $\mathbb{Z}_2$-symmetry would give $\sigma_0=-\sigma_L=6k\Mfive$ instead, which is the usual result \cite{RS1}).

\subsubsection{Matter solution}

As with the discrete theory, when we add a dynamical matter fluid to the system, we no longer have the freedom to solve the system for a conformally flat hypersurface metric, and need to keep things general. The metric ansatz that does the job for us is:
\begin{equation}\label{RS1_continuum}
    \dx{s}^2 = -c^2(t,y) \dx{t}^2 + a^2(t,y) \eta_{ij}\dx{x}^i\dx{x}^j + b^2(t,y) \dx{y}^2 \; ,
\end{equation}
where now it proves more convient to work with the conventional scale factors rather than warp factors. With this choice of metric, we get our 5D Einstein equations, of which there are four:
\begin{align}\label{G00}
    G_{00} &= 3\left(\frac{\dot{a}^2}{a^2}+\frac{\dot{a}}{a}\frac{\dot{b}}{b}\right) + 3\frac{c^2}{b^2} \left(\frac{a'}{a}\frac{b'}{b} -\frac{a''}{a}-\frac{a^{\prime2}}{a^2}\right) = \kappa^2 T_{00}
    \\
    \begin{split}\label{Gij}
    G_{ij} &= \frac{a^2}{c^2}\eta_{ij} \left(-\frac{\dot{a}^2}{a^2}+2\frac{\dot{a}}{a}\frac{\dot{c}}{c}-2\frac{\dot{a}}{a}\frac{\dot{b}}{b}+\frac{\dot{b}}{b}\frac{\dot{c}}{c}-2\frac{\ddot{a}}{a}-\frac{\ddot{b}}{b}\right) \\&+ \frac{a^2}{b^2}\eta_{ij} \left(\frac{a^{\prime2}}{a^2}-2\frac{a^\prime}{a}\frac{b^\prime}{b}+2\frac{a'}{a}\frac{c'}{c}-\frac{b'}{b}\frac{c'}{c}+2\frac{a''}{a}+\frac{c''}{c}\right) = \kappa^2 T_{ij}
    \end{split}
    \\
    G_{05} &= 3\left( \frac{\dot{a}}{a}\frac{c^\prime}{c}+\frac{a^\prime}{a}\frac{\dot{b}}{b}-\frac{\dot{a}'}{a}\right) = \kappa^2 T_{05}\label{G05}
    \\
    G_{55} &= 3\left( \frac{a^{\prime2}}{a^2}+\frac{a'}{a}\frac{c'}{c}\right) + 3\frac{b^2}{c^2}\left(\frac{\dot{a}}{a}\frac{\dot{c}}{c}-\frac{\dot{a}^2}{a^2}-\frac{\ddot{a}}{a}\right) = \kappa^2 T_{55}\label{G55} \; .
\end{align}
With bulk energy momentum tensor $T_{MN} = -2\Lambda_5 g_{MN}$, we immediately see that $G_{05}=0$. Remarkably, using this equation for $G_{05}$, one can show that any set of functions $a,b$ and $c$ which satisfy both $G_{05}=0$ and
\begin{equation}\label{LocalBulkEqn}
    \left(\frac{\dot{a}}{ac}\right)^2 = \left(\frac{a'}{ab}\right)^2 + \frac{\kappa^2\Lambda_5}{3} + \frac{C}{a^4} \; .
\end{equation}
will solve \emph{all} of Einstein's equations, locally in the bulk \cite{Full_EFEs_brane}. The term scaling as $a^{-4}$ is the so-called `dark radiation' term, which arises as a result of bulk Weyl curvature when the bulk is not AdS but AdS-Schwarzchild \cite{GeometryofBraneWorld}, so we are safe to set the constant $C$, which is proportional to the mass of the bulk black hole, to be 0.

If the matter on the brane (say, at $y=0$) is of perfect fluid form i.e $S_{\mu\nu}$ is given by Eq. \eqref{perfect_fluid_Tmunu}, then the boundary equation Eq. \eqref{Boundary} allows us to substitute in for $\left.(a'/ab)^2\right\vert_0$ to obtain a modified Friedmann equation for the dynamics on the brane\footnote{We could have also obtained this equation by using the Gauss-Codazzi equations \cite{Wald} to directly project the 5D Einstein tensor onto the brane at $y=0$.}:
\begin{equation}\label{RS_Friedmann}
    \left(\frac{\dot{a}_0}{a_0}\right)^2 = \frac{8\pi G}{3}\rho_0 \left(1+\frac{\rho_0}{2\sigma_0}\right)+\frac{\Lambda_4}{3}+\frac{C}{a_0^4} \; ,
\end{equation}
where Newton's constant and the 4D effective cosmological constant are, respectively, in terms of the 5D parameters, $8\pi G = 2\kappa^4\sigma/3$ and $\Lambda_4=\kappa^2(\Lambda_5 + \kappa^2\sigma^2/3)$. Again these are adapted from the standard RS identifications to our lack of $\mathbb{Z}_2$-symmetry -- though the dynamics remain the same (since $\rho_0$ also is affected by the $\mathbb{Z}_2$ factor). In particular, at early times we get a modification to the usual Friedmannn equation by a $\rho^2$ term, but since $\rho$ decays, at late times the evolution is as in standard FRW cosmology.

Since we have that the energy-momentum tensor on the brane is conserved, $\rho_0$ just follows the usual fluid equation:
\begin{equation}
    \dot{\rho_0} + 3\frac{\dot{a}_0}{a_0} (\rho_0 + p_0) = 0 \; ,
\end{equation}
so we have all the information we need to solve for the dynamics.

We can also extend the solution on the brane to the whole bulk, provided we take $b=1$ for all time (with some suitable radion stabilisation mechanism to allow for this, see e.g. \cite{Radion_stab}). The procedure is outlined in \cite{Full_EFEs_brane}, but here we will simply state the solution, which reads (when $C=0$):
\begin{align}
    a(t,y) &= a_0(t) (\cosh{ky} - \eta \sinh{k\abs{y}})
    \\
    c(t,y) &= \frac{\dot{a}(t,y)}{\dot{a}_0(t)} = \cosh{ky} - \tilde{\eta} \sinh{k\abs{y}}\label{lapse} \; ,
\end{align}
where $k$ is as in the static solution, and
\begin{equation}
    \eta = 1 + \frac{\rho_0}{\sigma} \; , \;\;\;\;\; \tilde{\eta} = \eta + \frac{\dot{\rho}_0a_0}{\dot{a}_0} \; .
\end{equation}

The equation \eqref{lapse} for the lapse in terms of the scale factor is a direct result of the Einstein equation $G_{05}=0$, which tells us that there is no flow of energy along the \nth{5} dimension. In the discrete theory, the lapses have the same form -- see Eq. \eqref{lapses_discrete} -- but there it is due to the Bianchi constraint on the $W$-tensor interaction terms, which tells us that there can be no energy flow across the sites of the clockwork. Clearly there is some correspondence between these two ideas -- more on this very shortly, but for now we continue on.

\section{Example Models}\label{Sec:Examples}

The upshot to all of the work in the previous Section is that in RS1 we have a fully solvable system in the continuum limit which we can use as a consistency check for our discrete clockwork gravity formalism. We shall show now explicitly how the discrete results map exactly to the RS results after taking the appropriate limit, and then go on to examine the solutions to the clockwork equations for the gravitational version of the toy quartic model introduced in Section \ref{Sec:QCS}.

\subsection{Randall-Sundrum (Discrete)}\label{Sec:RS}

Here we shall use the formalism developed in Section \ref{Sec:Discrete} to reproduce the results of Section \ref{Sec:Continuum} in the continuum limit. Recall that RS1 is the simplest possible continuum theory, with the continuum $\alpha$ parameters, which determine the size of the extrinsic curvature contributions to the bulk action, all set to 0. Using the dictionary of Eqs. \eqref{Tz}-\eqref{Tiii} for going between the discrete potential coefficients $T_{ijkl}$ and the continuum parameters, we see that the RS1 case corresponds to following choice of clockwork couplings (setting all discrete $\alpha_n=1$):
\begin{equation}\label{RS_betas}
    24\beta_0 = -\frac{6M^3_{(5)}}{\delta y}
    \; , \;\;\;
    24\beta_1 = \frac{3M^3_{(5)}}{\delta y}
    \; , \;\;\;
    24\beta_2 = -\frac{M^3_{(5)}}{\delta y}
    \; , \;\;\;
    \beta_3 = 0
    \; , \;\;\;
    24\beta_4 = 2\Lambda_5 \delta y \; .
\end{equation}
This choice for the $\beta$'s completely specifies the discrete theory.

We shall consider the static and dynamical solutions in turn.

\subsubsection{Static vacuum solution}

Comparison of our $H=0$ clockwork vacuum solution from Section \ref{Sec:Discrete} with the RS1 warped metric Eq. \eqref{RS1_static} tells us that we should identify $q=e^{k\delta y}$, so that we have the dictionary for going between clockwork and RS1 parameters:
\begin{equation}
    q = e^{k\delta y} \; , \;\;\;\;\; M_{(4)}^2 = \Mfive\delta y \; , \;\;\;\;\; L = (N-1)\delta y \; .
\end{equation}
The continuum limit is achieved, as we said, by taking the limit $N\rightarrow\infty$ and $\delta y\rightarrow 0$ while keeping $L=(N-1)\delta y$ fixed. In terms of the clockwork parameters, this is equivalent to taking $N\rightarrow\infty$, $q\rightarrow 1$ and $M_{(4)}^2\rightarrow 0$ while keeping $(N-1)M_{(4)}^2$ and $(N-1)\ln{q}$ fixed.

Substituting the RS $\beta$'s into the cosmological-constant-modified vacuum condition Eq. \eqref{vacuum_condition_sigmas}, and then taking the appropriate limit described above, to first order in $\delta y$ we find that in order for the vacuum condition to be satisfied (in the bulk) we require:
\begin{equation}\label{RSdiscretek}
    -\Lambda_5 = 3k^2\Mfive + 6k^3\Mfive\delta y \; ,
\end{equation}
and so we recover $k^2=-\Lambda_5/3\Mfive$, as we wanted, once we send $\delta y \rightarrow 0$.

The brane tensions, also to first order in $\delta y$, and utilising the above Eq. \eqref{RSdiscretek}, are:
\begin{align}
    \sigma_1 &= 3kM^3_{(5)} + \frac32 k^2 M^3_{(5)}\delta y
    \\
    \sigma_2 &= -3kM^3_{(5)} - \frac32 k^2 M^3_{(5)}\delta y \; ,
\end{align}
also in agreement with our continuum solution.

The zero-mode coupling Eq. \eqref{zero_mode_coupling}, in this limit, becomes:
\begin{equation}
    M^2_{\text{Pl}} = \frac{1}{2k}\left(e^{2kL}-1\right)\Mfive \; ,
\end{equation}
which again is the usual association one would make between the 4D and 5D Planck masses in static RS1.

\subsubsection{Matter solution}

We now wish to add matter to the brane at $y=0$, corresponding to the first site of the clockwork. While in the context of a clockwork model we would normally couple to the end site, to engineer the smallest coupling to the zero-mode, brane cosmology literature typically puts matter on the $y=0$ brane, so we do this too here for comparative purposes. In Appendix \ref{App:RS_eqs}, we show explicitly that, in the continuum limit, our discrete Einstein equations from Section \ref{Sec:Discrete} become the continuum Einstein equations from Section \ref{Sec:Continuum}. To be more precise, for sites in the bulk, the discrete equations \eqref{Friedmann} and \eqref{Raychaudhuri} for $G^0_{\;0}$ and $G^i_{\;j}$ become the continuum $G^0_{\;0}$ and $G^i_{\;j}$ equations \eqref{G00} and \eqref{Gij}, respectively, while for the two end sites the discrete equations for $G^0_{\;0}$ and $G^i_{\;j}$ respectively become the continuum Gibbons-Hawking equations \eqref{Boundary} for $K^0_{\;0}$ and $K^i_{\;j}$, corresponding to the component of $\partial\mathcal{M}$ in question. The Bianchi constraint \eqref{W constraint} also maps directly to the continuum $G^0_{\;5}$ equation, solidifying the correspondence eluded to at the end of Section \ref{Sec:Continuum}.

However, since the discrete theory is 4 dimensional, we have no notion of the $G^5_{\;5}$ equation, which was important when solving the continuum equations. Therefore, we need to check that the solutions to the discrete Einstein equations coincide with the solutions to the continuum Einstein equations, in the appropriate limit.

To do this, we used Python to solve the discrete Friedmann equations \eqref{adot0} using the framework set out at the end of Section \ref{Sec:Discrete}, i.e. we set the lapse on the first site to 1 and evolve the one dynamical equation for $\dot{a}_0$ via \nth{4} order Runge-Kutta, ensuring that the algebraic conditions \eqref{fi_fj} are satisfied at every time step of the evolution. To simplify things numerically, we choose units such that the parameters $\Mfive=1$, $\Lambda_5=-1$, and the size of the \nth{5} dimension is also $L=1$. We use these parameters also in the continuum Friedmann equation on the brane, Eq. \eqref{RS_Friedmann}, and solve it using a standard numerical ODE integrator, the hope being that the solutions to the continuum and discrete equations match in the limit where $N\rightarrow\infty$ and $\delta y = L/(N-1) \rightarrow 0$.

\begin{figure}[h!]
\centering
\begin{subfigure}{0.495\textwidth}
    \includegraphics[width=\textwidth]{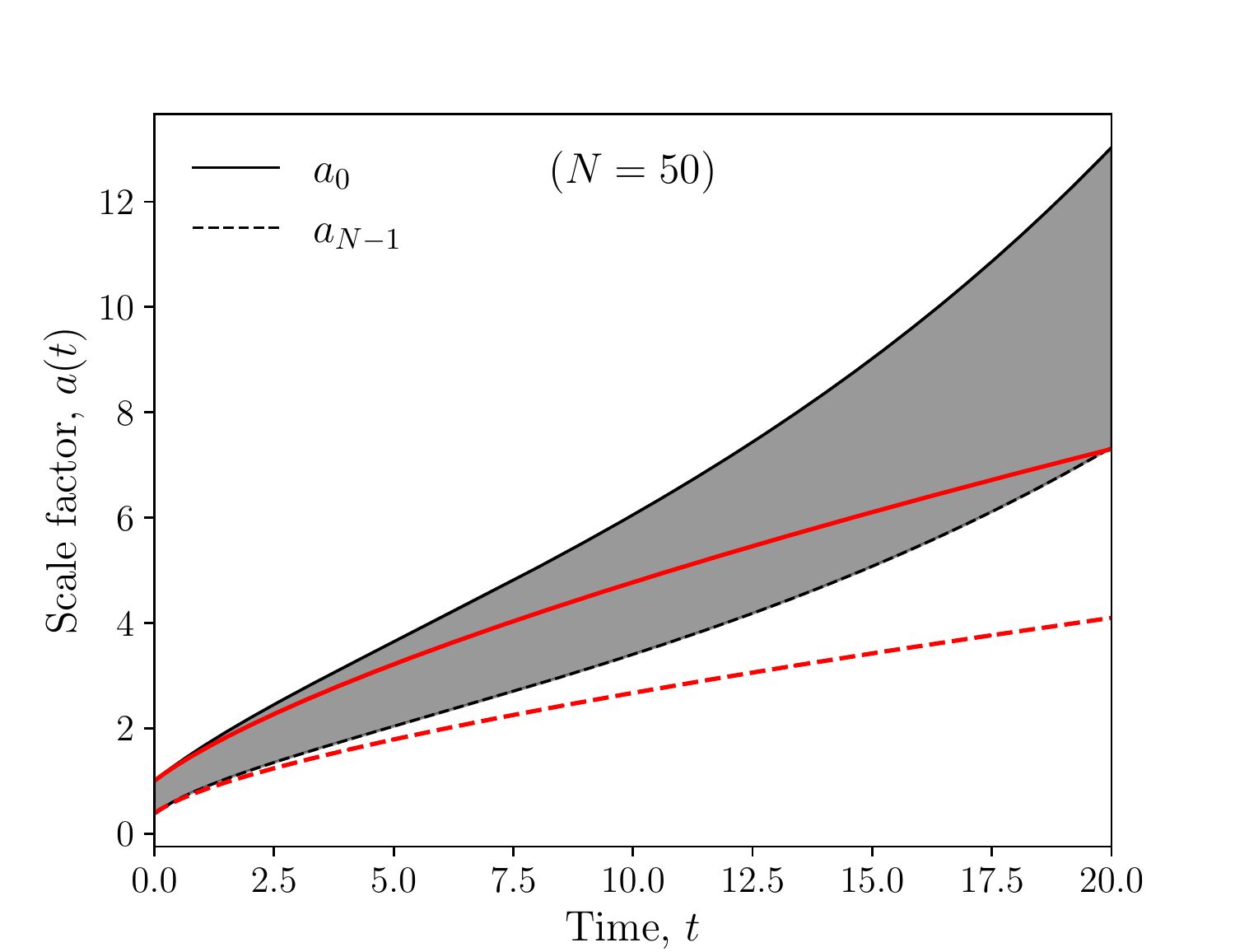}
    \caption{$N=50$ sites}
\end{subfigure}
\hfill
\begin{subfigure}{0.495\textwidth}
    \includegraphics[width=\textwidth]{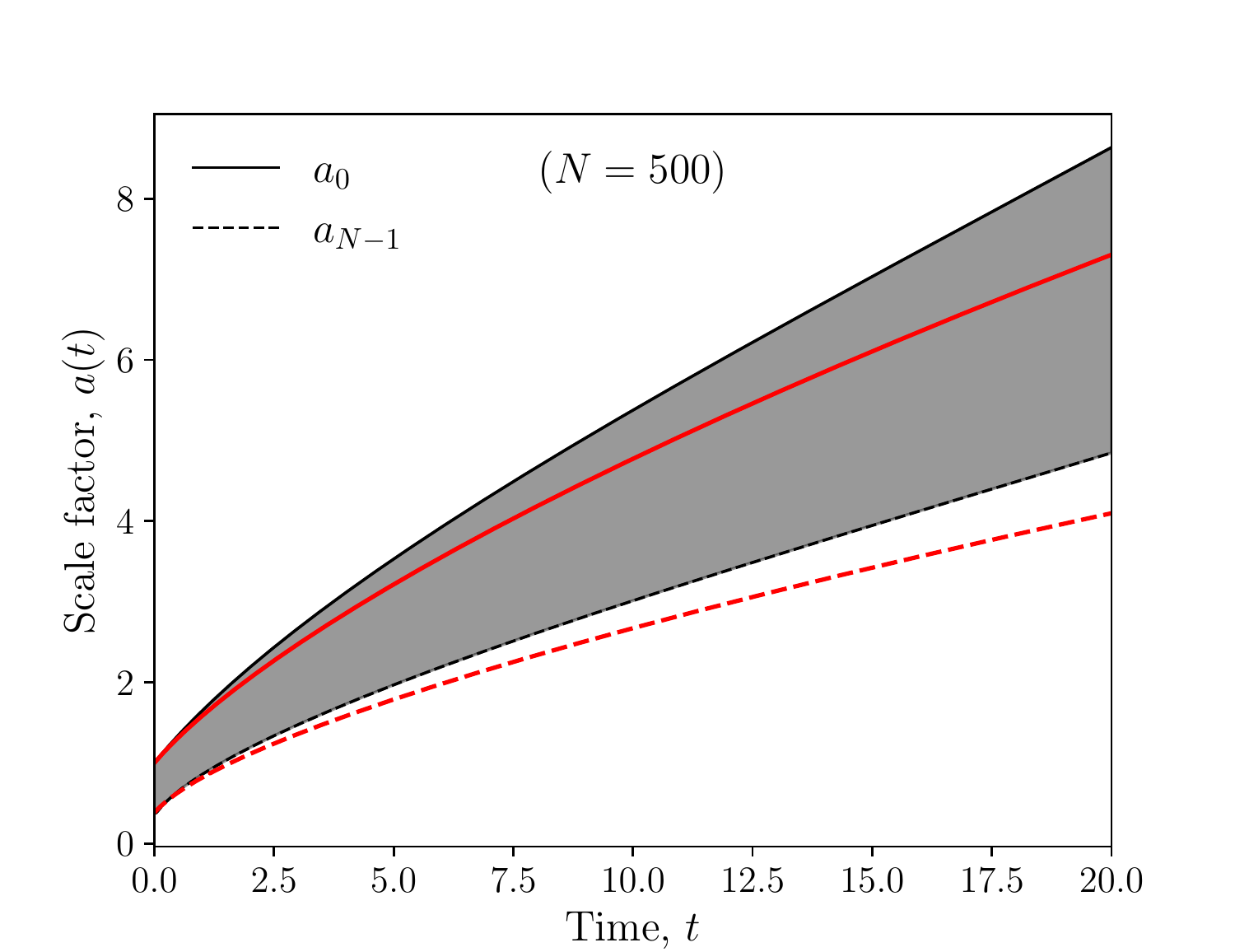}
    \caption{$N=500$ sites}
\end{subfigure}
\hfill
\begin{subfigure}{0.495\textwidth}
    \includegraphics[width=\textwidth]{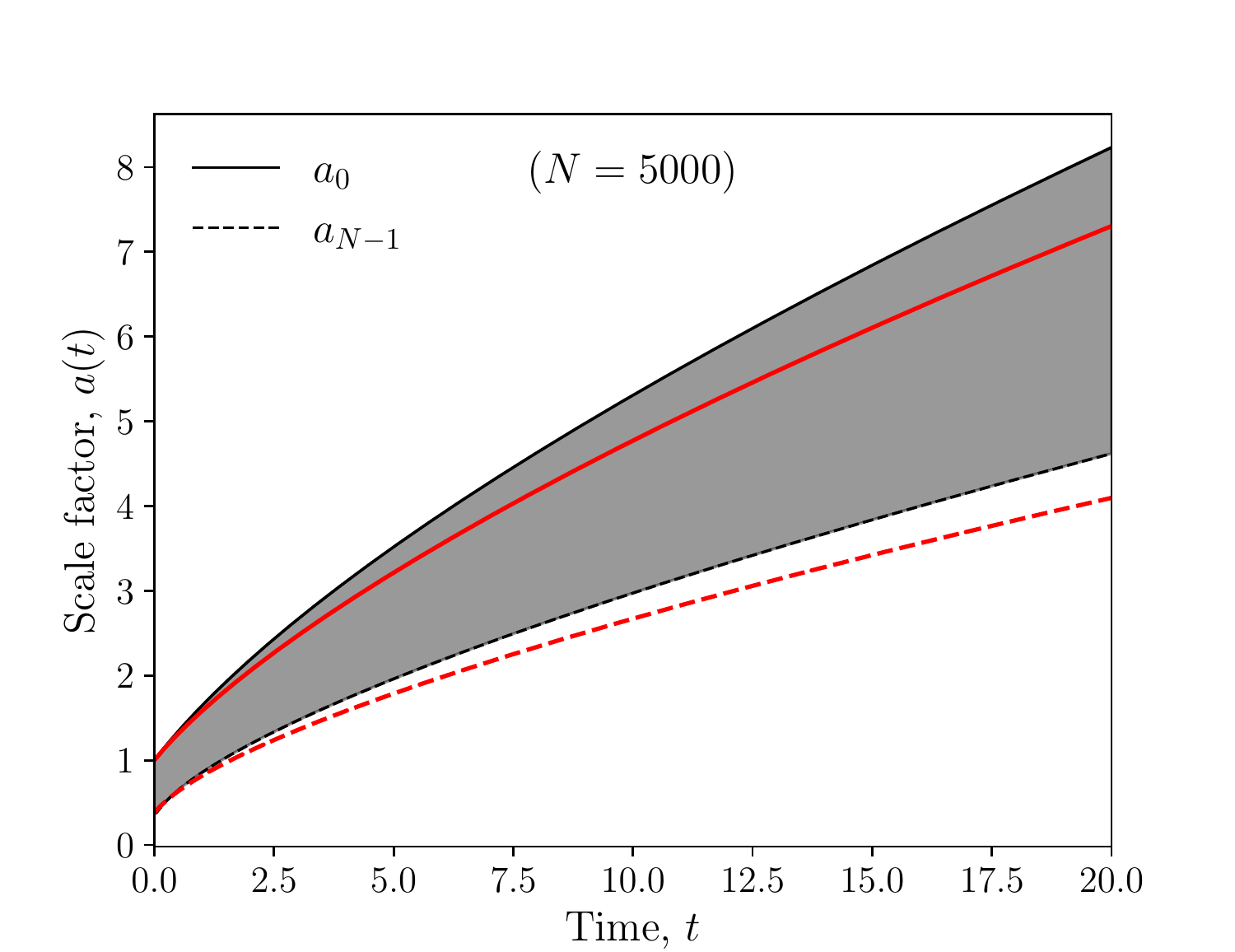}
    \caption{$N=5000$ sites}
\end{subfigure}
        
\caption{In black: solutions to the discrete Friedmann equations with potential couplings given by the RS $\beta$'s \eqref{RS_betas}; in red: solution to the continuum RS equations. Since the discrete system involves very many scale factors, whose solutions sit on top of one another sequentially, we only show explicitly the evolution of the first (solid line) and last (dashed line) sites and shade the region in between where the other scale factors lie. We see that as the number of sites is increased, the solution to the discrete equations indeed approaches that of the continuum -- this convergence is better seen in Fig \ref{Fig:RS_diffs}.}
\label{Fig:RS_sols}
\end{figure}

In Figure \ref{Fig:RS_sols} we show in black the solutions to the discrete equations when we add pressureless dust (i.e. with $\rho_0 = 1/a_0^3$ and $p_0=0$) to the first site of the clockwork, for $N=50, N=500$ and $N=5000$ sites, with the RS continuum solution overlaid in red with presureless dust on the brane at $y=0$.

\begin{figure}[h!]
\centering
    \includegraphics[width=0.8\textwidth]{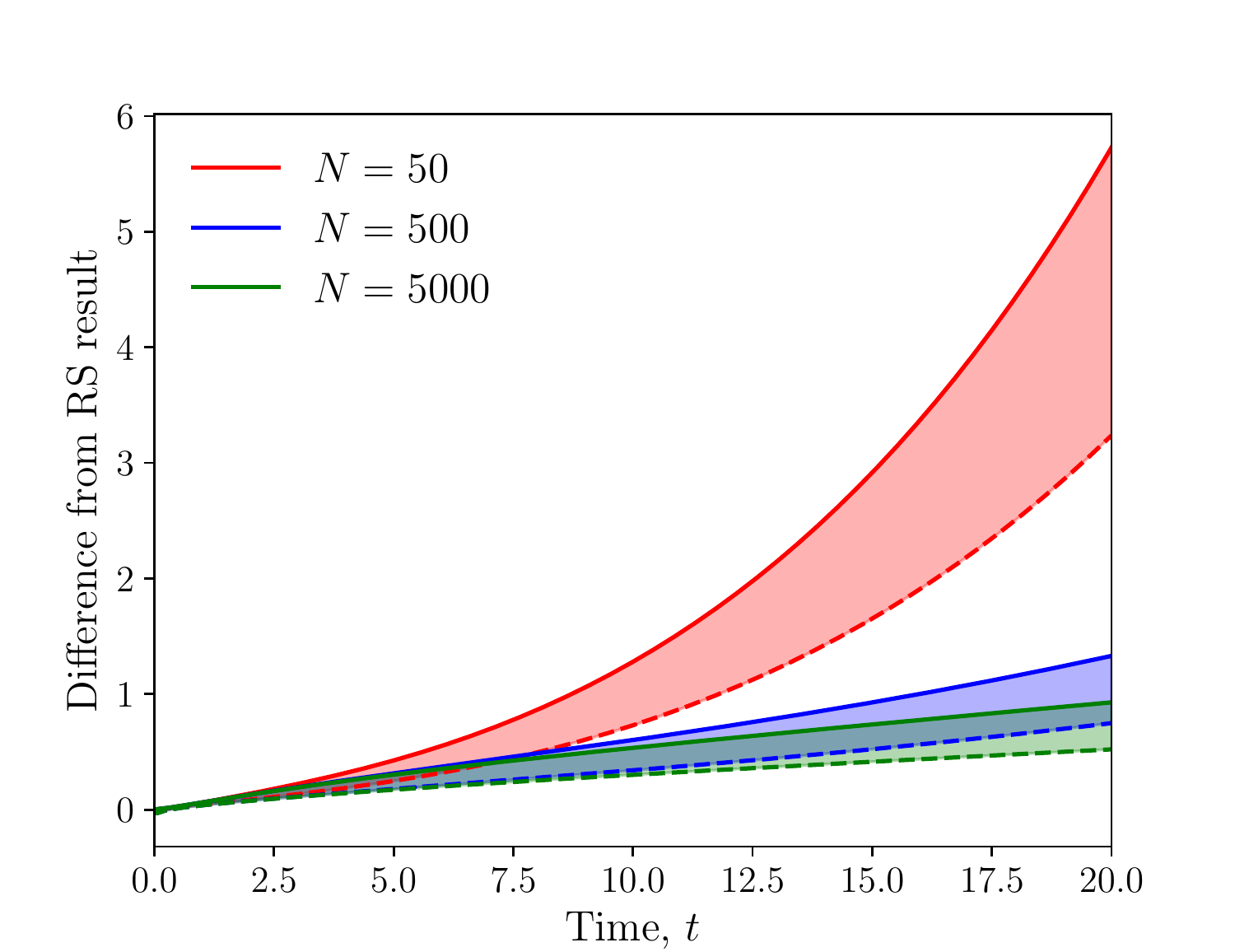}    
\caption{Difference between the continuum and discrete solutions displayed for the increasing number of sites as plotted in Fig \ref{Fig:RS_sols}, showing more clearly the convergence as we approach the continuum limit. Solid/dashed lines and the shaded regions have the same meaning as in Fig \ref{Fig:RS_sols}.}
\label{Fig:RS_diffs}
\end{figure}

Indeed, we see that the solutions begin to overlap as we get closer to the continuum limit, so we can be confident that our formalism works as intended, and that while we have no analogue for $G^5_{\;5}$ in the discrete system, the solutions are nevertheless equivalent. In Figure \ref{Fig:RS_diffs} we show the convergence as the number of sites increases explicitly.

We note that while the solution for small $N$ deviates markedly from the continuum solution, it is nevertheless a real, physical solution to the multi-gravity theory specified by the RS $\beta$'s. It is simply that this theory does not constitute a clockwork when one is away from the continuum limit. We can see this as follows: recall that the defining feature of clockwork gravity is that there exists a static vacuum solution where all metrics are conformally flat, and the corresponding conformal factors are asymmetrically distributed throughout the clockwork lattice, the canonical choice being $a_{i,0} = a_{0,0}/q^i$. For this solution to exist, the potential coefficients must satisfy the (in this case cosmological-constant-modified) vacuum condition \eqref{vacuum_condition_sigmas}, and for the RS $\beta$'s this is only true in the limit $\delta y \rightarrow 0$. We stress this point to make clear the fact that our formalism is entirely general, and can in principle be used to determine cosmological solutions to \emph{any} theory of gravity involving multiple pairwise interacting metrics -- in this case (i.e. for the theory with the RS $\beta$'s but away from the continuum limit) the solution is an accelerating one at late times. The novelty of clockwork gravity models, as a subclass of these general theories, is of course their potential to solve the hierarchy problem, so we pay them special attention.

As an example unstudied clockwork scenario, we now return to the quartic model introduced in Section \ref{Sec:QCS} for scalars, and construct its gravitational equivalent.

\subsection{Quartic clockwork gravity}\label{Sec:Giudice}

We introduced this model in Section \ref{Sec:QCS} as a simple $\deg=4$ clockwork theory with the nice feature of having the smallest possible hierarchy in the potential coefficients, with $\beta_m \propto q^{\pm 1}$ only, where $q>1$ is $\mathcal{O}(1)$. The potential coefficients specifying the theory are:
\begin{equation}\label{CW_betas}
    24\beta_0 = 6q^{-1}
    \; , \;\;\;
    24\beta_1 = -3
    \; , \;\;\;
    24\beta_2 = q
    \; , \;\;\;
    \beta_3 = 0
    \; , \;\;\;
    \beta_4 = 0 \; ,
\end{equation}
with all of the nonzero $\alpha_n = 1 \; \forall n\neq 0, N-1$. We rescale the $\beta$'s here by a factor 24 compared with those stated in Section \ref{Sec:QCS}, in order to force them in line with the Einstein equations. We are allowed to do this since we already showed that the $\beta$'s from Section \ref{Sec:QCS} indeed satisfy the vacuum condition (now without the need to include any $\sigma$'s on the end sites), so multiplying through by a common factor will not affect this. This time, in accordance with our clockwork philosophy, we will eventually place our matter on the end ($i=N-1$) site -- as this produces the greatest suppression of scales -- so we choose to fix the initial lapse and scale factor on that site to 1, which means we must have $c_i = \dot{a}_i/\dot{a}_{N-1}$ (see Eq. \eqref{lapses_discrete} and Appendix \ref{App:Bianchi constraint}).

With this choice of potential coefficients and lapse, the modified Friedmann equations take the form:
\begin{equation}
    3\Mfour \left(\frac{\dot{a}_{N-1}}{a_i}\right)^2 = \alpha_i \frac{6}{q} \left(1-\frac{q}{2}\frac{a_{i+1}}{a_i}\right)\left(1-q\frac{a_{i+1}}{a_i}\right) - 3\alpha_{i-1} \left(\frac{a_{i-1}}{a_i}\right)^3\left(1-q\frac{a_{i}}{a_{i-1}}\right) + \rho_i \; .
\end{equation}
Written in this form, the presence of the $H=0$ clockwork vacuum is manifest -- if $\rho_i = 0$ on all sites then the RHS vanishes when $a_{i+1,0}=a_{i,0}/q$.  In general we cannot find analytic forms for the dS vacua for more than 2 sites, as increasing the number of sites increases the order of the polynomial equation, Eq. \eqref{dS vacua}, which one must solve to determine the conformal factors. Indeed, we find numerically that the number of distinct (i.e. with different values for $H$) physical dS vacua increases with the number of sites, although the high degree of nonlinearity in the system prevents us from determining the exact quantity.

To solve for the dynamics when we add pressureless dust to the $(N-1)$-th site, we follow our usual procedure and evolve the one dynamical equation for $\dot{a}_{N-1}$ while ensuring that the algebraic conditions \eqref{fi_fj} are satisfied at all time steps. Since the theory possesses multiple vacua, this process is numerically sensitive to initial data -- indeed, if we set up the initial set of scale factors we feed into Python to be close to one of the dS vacua, then the addition of matter essentially acts as a perturbation to the system which quickly dilutes away (scaling as $a^{-3}$) as the system returns to the corresponding vacuum. However, if we choose initial scale factors close to the clockwork $H=0$ vacuum, then we do get interesting cosmological evolution. 

\begin{figure}[h!]
\centering
    \includegraphics[width=\textwidth]{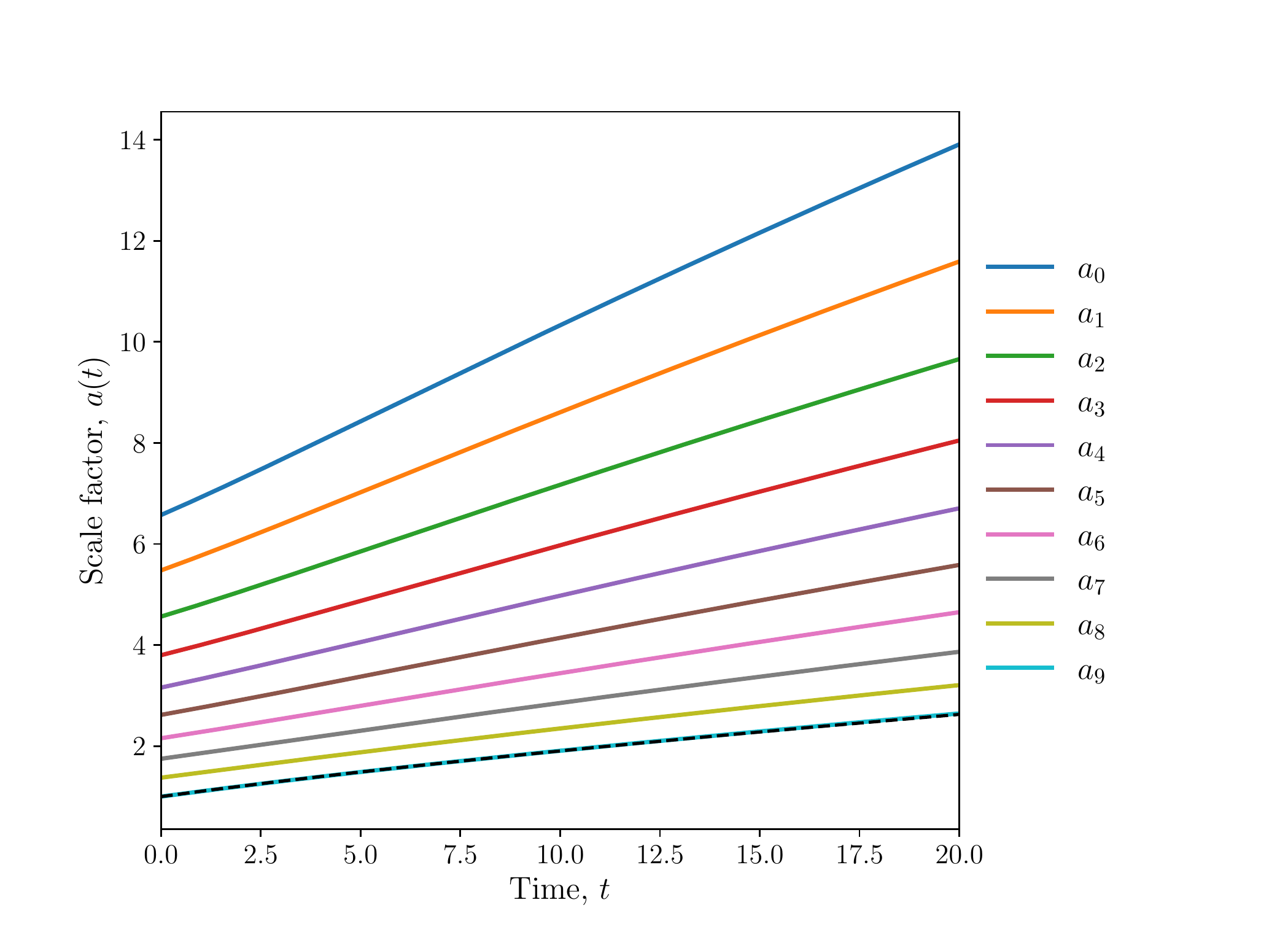}    
\caption{Solutions to the discrete Friedmann equations for $N=10$ sites with potential couplings given by the set of $\beta$'s \eqref{CW_betas}. The dashed black line is the effective evolution of the clockwork zero-mode (i.e. the solution to Eq. \eqref{zero_mode_evolution}) which we see matches almost exactly the evolution of $a_{N-1}$.}
\label{Fig:Cosmo_QCS}
\end{figure}

In Figure \ref{Fig:Cosmo_QCS}, the background evolution of the scale factors, with presureless dust on the end site, is displayed for $N=10$ sites, starting off at the clockwork vacuum, and for simplicity taking the parameters as $\Mfour=1/3$, $q=1.2$, and $\rho_{N-1}=1/a_{N-1}^3$. We are free to make these choices for $\Mfour$ and $\rho_{N-1}$ as these are essentially unit choices which correspond to a rescaling of $a_{N-1}$, so long as one also rescales the other scale factors in a consistent manner (taken care of by the algebraic equations). For comparison, we also plot in dashed black the effective evolution of a Universe whose gravity is described by GR with only a single FRW metric, with effective Planck mass $M_{\text{eff}}$ given by the zero-mode coupling Eq. \eqref{zero_mode_coupling}. That is, the solution to the standard Friedmann equation:
\begin{equation}\label{zero_mode_evolution}
    3M_{\text{eff}}^2 \left(\frac{\dot{a}}{a}\right)^2 = \rho \; ,
\end{equation}
with $\rho = 1/a^3$ as before.

We see that the system is solved by a bunch of successive scale factors undergoing decelerated expansions, but more importantly that the dynamics of the metric to which matter couples minimally are completely dominated by the contribution of the zero-mode. For the parameter choices given above -- numerically it appears that the following statement depends on the value of $q$, but it is certainly true here -- the effect of the clockwork interactions is such that, as far as the matter on the end site is concerned, the evolution is equivalent to that of a single copy of GR, whose strength is characterised by the Planck mass of the zero-mode (with some small corrections from the massive modes). This Planck mass, as we have seen, can be made much larger than the fundamental scale $M_{(4)}$ depending on one's choice for $q$ and $N$ (keeping in mind that we would like to avoid large hierarchies in the potential coefficients), and so in principle we see the potential route to solving the hierarchy problem -- in \cite{ClockworkGrav}, the authors showed that with $M_{(4)}\sim\mathcal{O}(\text{TeV})$ one can generate a Planck scale coupling to the zero-mode with $q=4$ and $N=26$ sites. This result should not be surprising, of course, since we are looking for cosmological solutions, and it is clear that the massless mode should dominate over large distance scales. The mass gap to the heavier modes is in general dependent on the choice of potential, but typically the mass gap to the lightest massive mode is roughly of order $qM_{(4)}$\footnote{Remember from Section \ref{Sec:Intro}, masses are products of scales ($M_{(4)}$) with couplings ($q$).}, with the heavier modes distributed exponentially above this. The strengths of their couplings to matter, determined by the components of the orthogonal matrix in Eq. \eqref{massive_modes}, are also generally model-dependent; for this model we find that the heaviest modes are even more weakly coupled than the zero-mode, which is favourable experimentally as their effects will not show up until above the Planck scale.

Of course, as mentioned at the end of the original clockwork gravity paper \cite{ClockworkGrav}, we still cannot yet present this as a robust solution to the hierarchy problem without considering radiative corrections to the potential. Graviton loops, for example, could generate non-nearest neighbour interactions, which would resurrect the undesirable Boulware-Deser ghost \cite{BD_Ghosts,Cycles}. Matter loops would not have this problem, since we chose to couple to only one site. We would naively expect that loops involving the graviton zero-mode would be safe due to the surviving diffeomorphism invariance, though loops involving the heavier modes could prove more dangerous, and a fuller analysis is required. Indeed, this discussion speaks to a more general question regarding the radiative stability of ghost-free multigravity theories (see \cite{drgt_corrections,bigravity_corrections} for work regarding this in the context of dRGT massive gravity and bigravity, to 1-loop level), which is outside the scope of the present work but important for future investigations.

\section{Conclusion}\label{Sec:Conclusion}

To summarise, the clockwork mechanism provides an efficient means to generate exponentially suppressed couplings from a fundamental theory containing only $\mathcal{O}(1)$ parameters, and as a result obtain exponentially large interaction scales without new physics appearing at these large energies. Applied to gravitational physics, one can naively solve the electroweak hierarchy problem through a higher order generalisation of the standard clockwork mechanism, involving nearest-neighbour interactions in the ghost free multi-gravity theory, where although the $N$ copies of diffeomorphism invariance are broken to the diagonal subgroup, an exponential suppression of the coupling to the graviton zero-mode is nevertheless achieved thanks to an asymmetric distribution of conformal factors in the background vacua.

In this work, we have done two main things. Firstly: we have developed a formalism by which one is able to construct potentials for which the general multi-gravity theory necessarily possesses the desired clockwork vacuum as a solution. To achieve this, we started with a general theory with nearest-neighbour interactions and used symmetric polynomials to constrain the choices of coefficients which are allowed in the potential. As an added bonus, we also get the matrix encoding the masses of the higher mass modes for free, and we used the formalism to reproduce some standard results in the clockwork literature, as well as to introduce a new model which has the nice feature of possessing the smallest possible hierarchy between the parameters of the fundamental theory.

Secondly: we derived the Einstein equations of the general ghost free multi-gravity theory, using the results from the symmetric polynomial formalism regarding the potential coefficients to aid in making them tractable. With the help of energy conservation considerations, we are able to solve the equations for a bunch of pairwise interacting FRW metrics whose lapse and scale factor are site-dependent. In vacuum, we can do this analytically, and find that there are a number of deSitter vacua where all scale factors evolve exponentially with time and every site shares the same Hubble parameter, the number of such solutions being in general dependent on the number of sites and a choice for the potential coefficients. In particular, if one imposes that there should be an asymmetrically distributed static vacuum solution with $H=0$, as required by the clockwork, then one recovers the condition on the allowed potential coefficients from the symmetric polynomials. When we add a minimally coupled perfect fluid to one of the sites, we must solve the equations numerically, and we provide a means to do this in an entirely general manner which works for \emph{any} multi-gravity theory devoid of the Boulware-Deser ghost, although we focus ourselves on clockwork theories i.e. those with the desired vacuum structure. We solved the equations explicitly for the new model constructed from the symmetric polynomial machinery and found that there is a cosmological solution where the evolution of the scale factor, as seen by a minimally coupled observer on the end site, looks like a single copy of GR whose strength is characterised by the exponentially enhanced Planck mass of the zero-mode.

We also revisited the continuum limit of the ghost free multi-gravity theory, which has a natural interpretation as a braneworld model, and made some clarifications about the nature of the extra dimension. Namely, the extra dimension cannot live on $S_1/\mathbb{Z}_2$, as is usually the case in braneworld literature, but rather, it must live on an interval, and so one must invoke a Gibbons-Hawking term to deal properly with the system boundaries. We paid special attention to the original Randall-Sundrum model (RS1) as a special case of the continuum theory, using it as a consistency check for our work on the discrete theory, and found that as one approaches the continuum limit, the solutions to the corresponding discrete theory (which is only a clockwork in this limit, otherwise it is just some generic multi-gravity theory) do indeed match up with the continuum results.

However, we cannot present this as a full resolution to the hierarchy problem just yet, as we still need to compute the radiative corrections, including those arising from graviton loops. We also need to investigate more of the cosmological phenomenology associated to the theory, in regards to perturbations and structure growth, black holes, gravitational waves etc. Since we only studied the explicit FRW solutions to a single model, these phenomenological considerations could be highly model-dependent. We have, however, provided a general framework within which one can work if one wishes to study such interesting questions; we certainly intend to do so in the near future.

\section*{Acknowledgements}

We would like to thank A. Padilla, S.S. Mishra and S. Sevillano Muñoz for useful discussions. K.W. is supported by a UK Science and Technology Facilities Council studentship. P.M.S. and A.A. are supported by a STFC Consolidated Grant [Grant No. ST/T000732/1]. For the purpose of open access, the authors have applied a Creative Commons Attribution (CC BY) licence to any Author Accepted Manuscript version arising.

\vspace{3em}

\newpage
\bibliographystyle{JHEP}
\bibliography{clockwork_refs.bib}


\newpage
\appendix

\section{Satisfying the Bianchi constraint}\label{App:Bianchi constraint}

We would like to see what the Bianchi constraint, Eq. \eqref{W constraint}, looks like explicitly with our FRW+lapse ansatz \eqref{dyn_metric} for the gear metrics.

Recall that the Bianchi constraint is
\begin{equation}\label{W constraint app}
    \sum_{i=0}^{N-1} \abs{e^{(i)}}\nabla^{(i)}_{\mu}W^{(i)\mu}_{\;\;\;\;\;\;\nu} = 0 \; .
\end{equation}
The index structure is written slightly differently to Eq. \eqref{W constraint}, but we are free to write it in this way since we use the Levi-Civita connection on each site, which is metric-compatible.

The covariant divergence of the $i$-th $W$-tensor is
\begin{equation}\label{Covar_full}
    \nabla^{(i)}_\mu W^{(i)\mu}_{\;\;\;\;\;\;\nu} = \partial_\mu W^{(i)\mu}_{\;\;\;\;\;\;\nu} + \Gamma^{(i)\mu}_{\;\;\;\;\mu\lambda}W^{(i)\lambda}_{\;\;\;\;\;\;\nu} - \Gamma^{(i)\lambda}_{\;\;\;\;\mu\nu}W^{(i)\mu}_{\;\;\;\;\;\;\lambda} \; .
\end{equation}
The only non-vanishing Christoffel symbols for our metrics are:
\begin{align}
    \Gamma^{(i)0}_{\;\;\;\;00} &= \frac{\dot{c}_i}{c_i}
    \\
    \Gamma^{(i)0}_{\;\;\;\;jk} &= \frac{a_i\dot{a}_i}{c_i^2}\eta_{jk}
    \\
    \Gamma^{(i)j}_{\;\;\;\;k0} &= \frac{\dot{a}_i}{a_i}\delta^j_k \; ,
\end{align}
and the only non-vanishing components of the $W$-tensors are $W^{(i)0}_{\;\;\;\;\;\;0}$ and $W^{(i)j}_{\;\;\;\;\;\;k}$, given respectively by Eqs. \eqref{W00} and \eqref{Wij}. Substituting into Eq. \eqref{Covar_full}, we get
\begin{align}
    \nabla^{(i)}_\mu W^{(i)\mu}_{\;\;\;\;\;\;0} &= \partial_0 W^{(i)0}_{\;\;\;\;\;\;0} + 3\frac{\dot{a}_i}{a_i} W^{(i)0}_{\;\;\;\;\;\;0} - \frac{\dot{a}_i}{a_i} W^{(i)k}_{\;\;\;\;\;\;k} 
    \\
     \nabla^{(i)}_\mu W^{(i)\mu}_{\;\;\;\;\;\;j} &= 0 \; ,
\end{align}
so we only need to consider the $\nu=0$ component.

Substituting in our explicit expressions for the $W$-tensor components yields (after some enjoyable algebra) the following final expression for the covariant divergence:
\begin{equation}\label{cov_div}
\begin{split}
     \nabla^{(i)}_\mu W^{(i)\mu}_{\;\;\;\;\;\;0} = 3\times 24 &\bigg[ \alpha_i (\dot{a}_{i+1}a_i^{-1} - c_{i+1}\dot{a}_ia_i^{-1}c_i^{-1})(\beta_1 + 2\beta_2 a_{i+1}a_i^{-1} + \beta_3 a_{i+1}^2 a_i^{-2})
     \\
     &+ \alpha_{i-1} (\dot{a}_{i-1}a_i^{-1} - c_{i-1}\dot{a}_ia_i^{-1}c_i^{-1})(\beta_1 a_{i-1}^2 a_i^{-2} + 2\beta_2 a_{i-1}a_i^{-1} + \beta_3 ) \bigg] \; .
\end{split}
\end{equation}

Clearly, the Bianchi constraint will be automatically satisfied when \emph{all} of these covariant divergences vanish individually; in fact, we showed in Section \ref{Sec:Discrete} that this \emph{has} to be the situation when matter only couples to one site (or when there is no matter at all). Since we want \eqref{cov_div} to vanish for every site, it must in particular vanish on the boundary sites ($i=0,N-1$), where one of the $\alpha$'s is 0. This means that both terms inside the square brackets must vanish individually, which tells us that we must have (both the $(i+1)$ and $(i-1)$ term give the same result),
\begin{equation}\label{lapses}
    \frac{c_{i+1}}{c_i} = \frac{\dot{a}_{i+1}}{\dot{a}_i} \implies c_i = \frac{\dot{a}_i}{f(t)} \; .
\end{equation}
To determine what the function of time $f(t)$ must be, we need only use the freedom to rescale our coordinates to fix the lapse on one of the sites. For example, if we want to set $c_0 = 1$, then we have $c_1/c_0 = \dot{a}_1/\dot{a}_0$, and so $f(t)=\dot{a}_0$. If we wish to work in conformal time and set $c_0 = a_0$, then we get $f(t)=\dot{a}_0/a_0$. If we wish to fix the lapse on the other end, i.e. take $c_{N-1}=1$, then we get $f(t)=\dot{a}_{N-1}$.

If matter couples to more than one site, there are of course in principle other non-trivial solutions to Eq. \eqref{W constraint app}, involving more complicated cancellations across the sums depending on the specific choice of $\beta$ coefficients.  However, these will not help us to do anything about the unknown lapses in our Einstein equations, and we only place matter on one of the sites anyway, so we use the solution \eqref{lapses} going forward. In fact, in Appendix \ref{App:RS_eqs} we show that the situation in Eq. \eqref{lapses} is the discrete analogue of the continuum $G^0_{\;5}$ Einstein equation (see Section \ref{Sec:Continuum}), so it is probably a good choice to make.

\section{RS continuum equations from the discrete equations}\label{App:RS_eqs}

We wish to show that, for the RS $\beta$'s, when we take the continuum limit, the discrete Einstein equations for the bulk scale factors become the corresponding 5D Einstein equations, the discrete equations for the boundary scale factors become the Gibbons-Hawking equations, and the Bianchi constraint becomes the $G^0_{\;5}$ equation.

Starting with $G^{0}_{\;0}$ for one of the bulk equations, and substituting in the RS $\beta$'s, we have explicitly:
\begin{equation}
\begin{split}
    -3\Mfour \left(\frac{\dot{a}_i}{a_ic_i}\right)^2 + \Bigg[ &\left( -\frac{6\Mfive}{\delta y} + \frac{9\Mfive}{\delta y}a_{i+1}a_i^{-1} - \frac{3\Mfive}{\delta y}a_{i+1}^2a_i^{-2}  \right)
    \\
    &+ \left(\frac{3\Mfive}{\delta y}a_{i-1}^3a_i^{-3} - \frac{3\Mfive}{\delta y}a_{i-1}^2a_i^{-2} + 2\Lambda_5\delta y \right)\Bigg] = 0 \; ,
\end{split}
\end{equation}
and we can Taylor expand the nearest neighbour scale factors to second order in $\delta y$:
\begin{align}
    a_{i-1} &= a_i - \delta y a_i' + \frac12 \delta y^2 a_i''
    \\
    a_{i+1} &= a_i + \delta y a_i' + \frac12 \delta y^2 a_i'' \; .
\end{align}
From now on we will suppress the $(i)$ indices for brevity. Substituting in the expansions, the first term in brackets becomes, to first order in $\delta y$,
\begin{equation}
    \left( \text{Bracket 1}  \right) \simeq 3\Mfive\frac{a^\prime}{a} - 3\Mfive \delta y \frac{a^{\prime2}}{a^2} + \frac32 \Mfive\delta y \frac{a''}{a} \; ,
\end{equation}
and the second term becomes:
\begin{equation}
    \left(\text{Bracket 2} \right) \simeq -3\Mfive\frac{a^\prime}{a} + 6\Mfive \delta y \frac{a^{\prime2}}{a^2} + \frac32 \Mfive\delta y \frac{a''}{a} + 2\Lambda_5\delta y \; .
\end{equation}
Adding the two together, and taking $\delta y\rightarrow 0$, we get the continuum equation:
\begin{equation}
    3\left(\frac{\dot{a}}{ac}\right)^2 - 3\left(\frac{a^{\prime2}}{a^2} + \frac{a''}{a}\right) = \frac{2\Lambda_5}{\Mfive} \; ,
\end{equation}
which is exactly $G^0_{\;0} = \kappa^2 T^0_{\;0}$.

On the boundaries, only one of the brackets is present, and the other is compensated for by a $(\sigma+\rho)$-type term. At $y=0$, we get
\begin{equation}
    -3\Mfive\delta y \left(\frac{\dot{a}_0}{a_0c_0}\right)^2 + 3\Mfive\frac{a^\prime}{a} - 3\Mfive \delta y \frac{a^{\prime2}}{a^2} + \frac32 \Mfive\delta y \frac{a''}{a} + (\sigma_0 +\rho_0) = 0 \; ,
\end{equation}
which in the limit yields:
\begin{equation}
    \left.\frac{a'}{a}\right\vert_0 = -\frac{\sigma_0+\rho_0}{3\Mfive} \; ,
\end{equation}
which is, as expected, $K^0_{\;0} - Kh^0_{\;0} = -\kappa^2 S^0_{\;0}$.

On the opposite boundary, we get
\begin{equation}
    -3\Mfive\delta y \left(\frac{\dot{a}_{N-1}}{a_{N-1}c_{N-1}}\right)^2  -3\Mfive\frac{a^\prime}{a} + 6\Mfive \delta y \frac{a^{\prime2}}{a^2} + \frac32 \Mfive\delta y \frac{a''}{a} + 2\Lambda_5\delta y + (\sigma_{N-1} +\rho_{N-1}) = 0 \; ,
\end{equation}
which in the limit yields:
\begin{equation}
    \left.\frac{a'}{a}\right\vert_{N-1} = +\frac{\sigma_{N-1}+\rho_{N-1}}{3\Mfive} \; ,
\end{equation}
which is, as expected, $K^0_{\;0} - Kh^0_{\;0} = +\kappa^2 S^0_{\;0}$. All is well thus far.

Next, we move to the considerably more complicated $G^i_{\;j}$ equation. For this, we also need to Taylor expand the lapses, as $c_{i+1}$ and $c_{i-1}$ terms now appear in the $W$-tensor as well. Explicitly, the equations are:
\begin{equation}
    \begin{split}
        \frac{M_{(4)i}^2}{c_i^2} \left(\frac{\dot{a}^2_i}{a^2_i} + 2\frac{\ddot{a}_i}{a_i} - 2\frac{\dot{a}_i}{a_i}\frac{\dot{c}_i}{c_i}\right) \\= 24 \Biggl\{ &\biggl[  -\frac{6\Mfive}{\delta y} + \frac{3\Mfive}{\delta y}(c_{i+1}c_i^{-1} + 2a_{i+1}a_i^{-1}) \\&-\frac{\Mfive}{\delta y}(2c_{i+1}c_i^{-1}a_{i+1}a_i^{-1} + a_{i+1}^2a_i^{-2}) \biggr]
        \\
        &+ \biggl[ \frac{3\Mfive}{\delta y}(c_{i-1}c_i^{-1}a_{i-1}^2a_i^{-2}) \\&-\frac{\Mfive}{\delta y} (2c_{i-1}c_i^{-1}a_{i-1}a_i^{-1} + a_{i-1}^2a_i^{-2})
        +2\Lambda_5\delta y \biggr] \Bigg\} \; .
    \end{split}
\end{equation}
Performing our Taylor expansion on both $a$ and $c$, the first term in the square brackets becomes:
\begin{equation}
\begin{split}
    (\text{Sq. bracket 1}) &\simeq \Mfive\frac{c'}{c} + 2\Mfive\frac{a'}{a} + \Mfive\delta y \frac{a''}{a} + \frac12 \Mfive\delta y \frac{c''}{c} \\&- 2\Mfive\delta y \frac{a'}{a}\frac{c'}{c} - \Mfive\delta y\frac{a^{\prime2}}{a^2} \; ,
\end{split}
\end{equation}
and the second one is:
\begin{equation}
\begin{split}
    (\text{Sq. bracket 2}) &\simeq -\Mfive\frac{c'}{c} - 2\Mfive\frac{a'}{a} + \Mfive\delta y \frac{a''}{a} + \frac12 \Mfive\delta y \frac{c''}{c} \\&+ 4\Mfive\delta y \frac{a'}{a}\frac{c'}{c} + 2\Mfive\delta y\frac{a^{\prime2}}{a^2} + 2\Lambda_5\delta y \; .
\end{split}
\end{equation}
Summing and taking $\delta y\rightarrow 0$, we get our continuum equation:
\begin{equation}
    \frac{1}{c^2} \left(2\frac{\dot{a}}{a}\frac{\dot{c}}{c} -\frac{\dot{a}^2}{a^2}-2\frac{\ddot{a}}{a}\right) + \left(2\frac{a''}{a} + \frac{a^{\prime2}}{a^2} + 2\frac{a'}{a}\frac{c'}{c} + \frac{c''}{c} \right) = -\frac{2\Lambda_5}{\Mfive} \; ,
\end{equation}
which is, of course, $G^i_{\;j}=\kappa^2 T^i_{\;j}$.

On the boundaries, we get the following at $y=0$,
\begin{equation}
    \left.\left(-2\frac{a'}{a} - \frac{c'}{c}\right)\right\vert_0 = -\frac{p_0-\sigma_0}{\Mfive} \; ,
\end{equation}
which is precisely $K^i_{\;j} - Kh^i_{\;j}= -\kappa^2 S^i_{\;j}$. At the $y=L$ boundary, we get the respective minus sign accounting for the change in orientation. So, once again, all is well.

Finally, we come to the Bianchi constraint. For the RS $\beta$'s, Eq. \eqref{cov_div} for the vanishing of the individual covariant divergences of the $W$-tensors becomes:
\begin{equation}
\begin{split}
     3 &\Bigg[ \alpha_i \left(\dot{a}_{i+1}a_i^{-1} - c_{i+1}\dot{a}_ia_i^{-1}c_i^{-1}\right)\left(\frac{3\Mfive}{\delta y} - 2\frac{\Mfive}{\delta y} a_{i+1}a_i^{-1}\right)
     \\
     &+ \alpha_{i-1} \left(\dot{a}_{i-1}a_i^{-1} - c_{i-1}\dot{a}_ia_i^{-1}c_i^{-1}\right)\left(\frac{3\Mfive}{\delta y} a_{i-1}^2 a_i^{-2}  - 2\frac{\Mfive}{\delta y} a_{i-1}a_i^{-1} \right) \Bigg] = 0 \; .
\end{split}
\end{equation}
Taylor expanding to first order in $\delta y$,
\begin{equation}
     3\Bigg[ \alpha_i \left(\frac{\dot{a}'}{a} - \frac{\dot{a}}{a}\frac{c'}{c}\right)\delta y \left(\frac{\Mfive}{\delta y} + \mathcal{O}(1) \right)
     + \alpha_{i-1} \left(-\frac{\dot{a}'}{a} + \frac{\dot{a}}{a}\frac{c'}{c}\right)\delta y \left(\frac{\Mfive}{\delta y} + \mathcal{O}(1) \right) \Bigg] = 0 \; .
\end{equation}
Taking $\delta y \rightarrow 0$, we get the continuum equation:
\begin{equation}
    3\Mfive \left(\frac{\dot{a}'}{a} - \frac{\dot{a}}{a}\frac{c'}{c}\right) (\alpha_i - \alpha_{i-1}) = 0 \; .
\end{equation}
Obviously this is true in the bulk where all $\alpha$'s are equal, but it must also hold on the boundaries, where one of the $\alpha$'s is 0. Hence, we get (in complete analogy with the process in Appendix \ref{App:Bianchi constraint})
\begin{equation}
     3\Mfive \left(\frac{\dot{a}'}{a} - \frac{\dot{a}}{a}\frac{c'}{c}\right) = 0 \; ,
\end{equation}
which is precisely $G^0_{\;5}=0$ -- in other words, covariant conservation of the interactions between the gears in the discrete theory is the analogue of there being no flow of energy-momentum along the \nth{5} dimension.

\end{document}